\documentclass[a4paper,onecolumn,11pt,accepted=2025-12-07]{quantumarticle}
\pdfoutput=1

\usepackage{graphicx}
\usepackage[normalem]{ulem}
\usepackage{color,bm}
\usepackage[caption=false]{subfig}
\usepackage[numbers, sort&compress]{natbib}
\usepackage{latexsym}
\usepackage{array,multirow}
\usepackage{hyperref}

\usepackage[utf8]{inputenc}

\definecolor{xred}{rgb}{1,0,0}

\definecolor{xblue}{rgb}{0,0,1}
\def  \Blue#1{\textcolor{xblue}{#1}}
\definecolor{xgreen}{rgb}{0,1,0}
\def  \Green#1{\textcolor{xgreen}{#1}}

\begin{document}

\title{Multiple quantum exceptional, diabolical, and hybrid points in multimode bosonic systems:
I. Inherited and genuine singularities}

\author{Kishore Thapliyal}
\email{kishore.thapliyal@upol.cz}
\orcid{0000-0002-4477-6041}
\affiliation{Joint Laboratory of
Optics, Faculty of Science, Palack\'{y} University, Czech
Republic, 17. listopadu 12, 771~46 Olomouc, Czech Republic}

\author{Jan Pe\v{r}ina Jr.}
\email{jan.perina.jr@upol.cz} \affiliation{Joint Laboratory of
Optics, Faculty of Science, Palack\'{y} University, Czech
Republic, 17. listopadu 12, 771~46 Olomouc, Czech Republic}

\author{Grzegorz Chimczak}
\affiliation{Institute of Spintronics and Quantum Information,
Faculty of Physics, Adam Mickiewicz University, 61-614 Pozna\'{n},
Poland}

\author{Anna Kowalewska-Kud\l{}aszyk}
\affiliation{Institute of Spintronics and Quantum Information,
Faculty of Physics, Adam Mickiewicz University, 61-614 Pozna\'{n},
Poland}

\author{Adam Miranowicz}
\affiliation{Institute of Spintronics and Quantum Information,
Faculty of Physics, Adam Mickiewicz University, 61-614 Pozna\'{n},
Poland}

\begin{abstract}
The existence and degeneracies of quantum exceptional, diabolical,
and hybrid (i.e., diabolically degenerated exceptional)
singularities of simple bosonic systems composed of up to five
modes with damping and/or amplification are analyzed. Their
dynamics governed by quadratic non-Hermitian Hamiltonians is
followed using the Heisenberg-Langevin equations. Their
dynamical matrices generally exhibit specific structures that
allow for an effective reduction of their dimension by half. This
facilitates analytical treatment and enables efficient spectral
analysis based on characteristic second-order diabolical
degeneracies. Conditions for the observation of inherited quantum
hybrid points, observed directly in the dynamics of field
operators, having up to third-order exceptional and second-order
diabolical degeneracies are revealed. Surprisingly, exceptional
degeneracies of only second and third orders are revealed, even
though the systems with up to five modes are considered.
Exceptional and diabolical genuine points and their degeneracies
observed in the dynamics of second-order field-operator moments
are also analyzed.  {Each analyzed bosonic system exhibits its
own unique and complex dynamical behavior.}

\end{abstract}

\maketitle

\section{Introduction}

Non-Hermitian Hamiltonians had been for a long time considered not
being suitable for describing real physical systems. This opinion
has changed after the seminal work by Bender and
Boettcher~\cite{Bender1998} who showed that the non-Hermitian
Hamiltonians endowed with a parity and time symmetry
($\mathcal{PT}$-symmetry) exhibit real spectra in certain areas of
the system parameter space. This leads to the formulation of new
area of physics, i.e., non-Hermitian quantum
mechanics~\cite{Bender2003,Bender2007, El-Ganainy2018,Ashida2020},
which has already provided numerous models \cite{Mostafazadeh2003,
Mostafazadeh2004,Mostafazadeh2010,Znojil2008,Brody2014,Bagarello2016}
suitable for describing real quantum systems in many areas of
physics. Moreover, non-Hermitian Hamiltonians exhibit new
algebraic structures: It has been shown that, for certain values
of parameters, there occur spectral degeneracies accompanied by
degeneracies of the eigenvectors of a given Hamiltonian. Such
points in a system parameter space with exceptional degeneracies
(EDs) are called exceptional points (EPs) for which the dimension
of the corresponding Hilbert space is reduced. This has
interesting physical consequences and leads to new unexpected
physical effects. It allows to enhance the precision of
measurements of suitable physical
quantities~\cite{Chen2017,Liu2016,Feng2017,El-Ganainy2019,Parto2021}.
It also leads to enhanced nonlinear interactions
\cite{PerinaJr2019b,PerinaJr2019c}. For this reason,
$\mathcal{PT}$-symmetric systems of different kinds have been
found appealing in many areas of physics including: optical
waveguides~\cite{Turitsyna2017,Xu2018}, optical
lattices~\cite{Graefe2011, Miri2012,Ornigotti2014,Shui2019}, spin
lasers \cite{Drong2022}, optical coupled
structures~\cite{El-Ganainy2007,
Ramezani2010,Zyablovsky2014,Ogren2017,Naikoo2019}, coupled optical
microresonators~\cite{Peng2014,Peng2014a,Liu2016,Zhou2016,Arkhipov2019,Arkhipov2020},
quantum-electrodynamics circuits (QED)~\cite{Quijandria2018},
systems with complex potentials~\cite{Guo2009}, optomechanical
systems~\cite{Tchodimou2017,Wang2019}, photonics
molecules~\cite{El-Ganainy2014}, among others. Moreover, schemes
for engineering properties of EPs have been suggested (see, e.g.,
\cite{Minganti2022} and references therein).

Subsequent studies have revealed also other features observed in
$\mathcal{PT}$-symmetric systems in addition to EPs. For example,
the spectral degeneracies which are not accompanied by the
corresponding eigenvector degeneracies, were observed
\cite{Berry1984}. Such degeneracies do not usually lead to the
above discussed physical effects and so the corresponding points
in the system parameter space were named diabolical points (DPs).
Note that DPs, contrary to EPs, can also be observed in Hermitian
systems \cite{Berry1984}. As pointed out in~\cite{PerinaJr2022a},
it may happen that this DP with its diabolical degeneracy (DD)
occurs at an EP. In this case, `independent' (i.e. with different
eigenvectors) multiple spectral and eigenvector degeneracies are
found in systems and we refer to them as hybrid diabolic
exceptional points (HPs). Such systems then naturally exhibit a
physical behavior similar to that observed at EPs. Moreover, new
effects originating in diabolical degeneracy may arise. For
example, the system behavior when encircling an HP has been used
to construct a multi-mode optical switch \cite{Arkhipov2023a}.

Non-Hermitian $\mathcal{PT}$-symmetric optical bosonic systems are
especially interesting from the point of view of their behavior at
EPs. Their infinite-dimensional Hilbert space leads to numerous
manifestations of modifications of their dynamics at quantum EPs
(QEPs), i.e. EPs observed in quantum systems \cite{PerinaJr2019c},
including the effect of quantum jumps
\cite{Minganti2019,Minganti2020}. The system dynamics may be
followed either directly in the Hilbert space [the Liouville space
of statistical operators] or in the complementary space of
field-operator moments (FOMs) of all orders
\cite{Perina1991,PerinaJr1995}. The studies performed in the
moment space of field operators in multimode bosonic systems
described by non-Hermitian quadratic Hamiltonians revealed
different types of degeneracies related to
QEPs~\cite{Arkhipov2021,PerinaJr2022a}.  Moreover, they allowed to
sort QEPs into three classes: inherited QEPs, genuine QEPs and
induced QEPs. The dynamical equations for the mean values of field
operators indicated the presence of inherited QEPs and quantum HPs
(QHPs) \cite{PerinaJr2022a} that represent the core of the studied
unusual behavior. The presence of such inherited QEPs and QHPs
then implies the existence of genuine QEPs and QHPs
\cite{PerinaJr2022a} observed in the dynamics of higher-order
FOMs. With the increasing FOM order, the degeneracies of genuine
QEPs and QHPs increase. Moreover, similar or identical FOMs, as
being related by the field commutation relations, arise in the
formal construction of higher-order FOMs. Thus, we can also define
induced QEPs and QHPs \cite{PerinaJr2022a} that further enlarge
the multiplicity of the spectral degeneracies. However, as these
redundant FOMs share their time evolution with the FOMs
contributing to genuine QEPs and QHPs, they do not lead to
additional diversity of the system evolution. Thus, they are not
interesting from the point of view of the dynamics of FOMs of a
given order. This dynamics is fully characterized by the
corresponding genuine QEPs and QHPs. As the properties of genuine
QEPs and QHPs originate in those of the inherited QEPs and QHPs,
the analysis of the latter is crucial for the understanding of a
system evolution. For this reason, it is important to identify the
inherited QEPs and QHPs and their degeneracies in simple bosonic
systems formed by smaller numbers of bosonic modes. This analysis
may then be exploited for further studies of physical effects in
such systems.

The system composed of two mutually interacting modes, one being
damped and the other amplified, was already analyzed from this
point of view in Ref.~\cite{PerinaJr2022a}. This analysis may be
considered as the simplest building block useful for
investigations of more complex bosonic systems  {in which much
richer and diverse physical behavior is expected}. Here, we
consider systems composed of up to five modes in different
configurations promising for the observation of QEPs and QHPs.
Looking for inherited QEPs with higher-order EDs is the main goal
of our investigations. It is motivated by the fact that the higher
is the ED order, the more modified is the system dynamics at a
QEP. This then enhances the physical effects specific to QEPs like
improvement in measurement precision or enhancement of nonlinear
effects.

As, surprisingly, we have been able to reveal only up to the
third-order ED of QEPs in the analyzed systems, we continue our
analysis in Ref.~\cite{PerinaJr2025}, being the second part of
this paper, in which we pay attention to the spectral degeneracies
of non-Hermitian Hamiltonians observed only in specific subspaces
of the systems' Liouville spaces as well as non-Hermitian
Hamiltonians with unidirectional coupling.  {In particular,
systems with unidirectional coupling exhibit significantly
stronger non-Hermitian features compared to those analyzed here,
enabling the observation of QEPs and QHPs of arbitrary order.} In
Ref.~\cite{PerinaJr2025}, we also address numerical identification
of QEPs and QHPs, useful in the bosonic systems with higher
dimensions. We also extend the analysis of the genuine and induced
QEPs and QHPs by considering the FOMs of a general order.

 {At the end of Introduction, we would like to note that while
the existence of higher-order EPs has been discussed previously in
the literature, those analyses typically rely on semiclassical
models that neglect quantum fluctuations, that are necessary for
quantum consistency of the models. Moreover, many prior works
investigate the spectra of semiclassical non-Hermitian
Hamiltonians, which exhibit infinite (though countable) sets of
eigenfrequencies. In contrast, the physically relevant spectra are
those of the corresponding dynamical matrices governing the
evolution equations (e.g., the Heisenberg ones in the Heisenberg
picture), which involve only a small number of frequencies. It is
precisely these frequencies that determine the observable behavior
of the system. In semiclassical models, higher-order EPs can often
be identified with relative ease. However, the validity of such
models is generally restricted to specific conditions, such as
short evolution times or weak damping/amplification. This raises a
critical and largely unexplored question: What are the spectral
degeneracies in PT-symmetric bosonic systems under general
conditions, where fluctuating quantum forces - governed by
fluctuation-dissipation theorems - play a non-negligible role?}

The paper is organized as follows. A two-mode bosonic system with
unequal damping and/or amplification rates, as the simplest
considered model, is analyzed in Sec.~II. Section~III brings the
analysis of a related three-mode system in the linear
configuration. The corresponding generalized four-mode systems in
their linear and circular configurations are investigated in
Sec.~IV whereas the analysis of the five-mode systems in their
linear and pyramid configurations is found in Sec.~V. Section~VI
brings conclusions. In Appendix~\ref{AppA}, the eigenvalues and
eigenvectors of the general $ 2n\times 2n $ dynamical matrices are
summarized.  {The structure of the second-order FOMs for the
analyzed systems is detailed in the tables provided in
Appendix~\ref{AppB}.}

\section{Two-mode bosonic system: Basic building blocks}\label{sec2}

\begin{figure*}  
 \hspace{10mm}
 \begin{centering}
  \includegraphics[width=0.85\hsize]{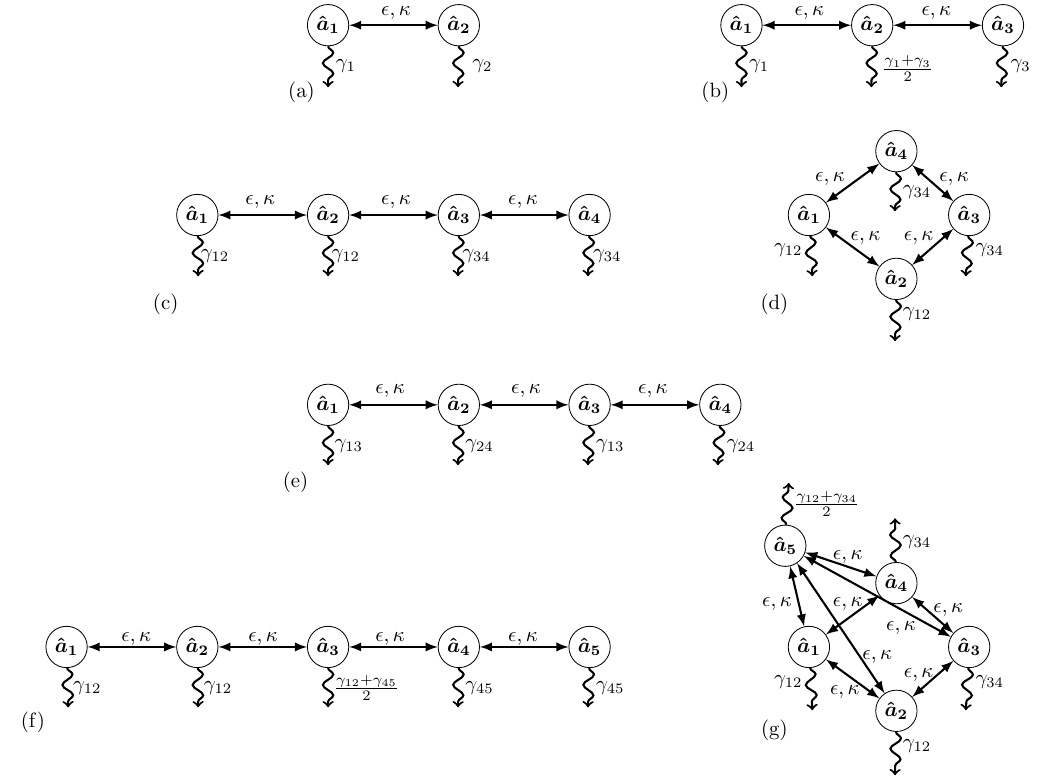}
 \end{centering}

 \caption{Schematic diagrams of the bosonic systems composed of (a) two, (b) three, (c---e) four, and (f,g) five modes
 with typical linear, circular, and pyramid configurations  {that exhibit quantum exceptional points (QEPs) and quantum hybrid points (QHPs) with
 various exceptional degeneracies (EDs) and diabolical
 degeneracies (DDs) observed in the dynamics of field-operator
 moments (FOMs) of different orders.} Strengths $ \epsilon $ and $ \kappa $
 characterize, respectively, the linear and nonlinear coupling
 between the modes, $ \gamma $, with subscripts indicating the mode number(s), are the
 damping or amplification rates, and
 annihilation operators $ \hat{a} $ identify the mode number via their subscripts,
 and $ \gamma_{jk} $ indicates that $ \gamma_j = \gamma_k $.}
\label{fig1}
\end{figure*}

We begin with the consideration of one of the simplest
$\mathcal{PT}$-symmetric bosonic systems that is composed of two
modes: one being attenuated and the other amplified [for the
scheme, see Fig.~\ref{fig1}(a)]. We note that even a one-mode
bosonic system may exhibit $\mathcal{PT}$-symmetric behavior, as
shown in Ref.~\cite{Arkhipov2023}. We also pay attention only to
the systems described by quadratic Hamiltonians that lead to
linear exactly-solvable Heisenberg equations. Though these
Hamiltonians lead to linear dynamical equations, they allow for
describing the nonlinear effect of photon-pair generation and
annihilation. This effect is commonly used in various quantum
optical systems to generate entangled \cite{Mandel1995} and
squeezed \cite{Luks1988,Dodonov2002} states of light.
$\mathcal{PT}$-symmetry restricts the form of the studied
non-Hermitian Hamiltonian such that the underlying dynamics is
described by the dynamical matrix composed of characteristic $
2\times 2 $ submatrices. They are also used to build the matrices
of more complex $\mathcal{PT}$-symmetric bosonic systems.  {The
characteristic block structure --- typical of bosonic systems with
quadratic Hamiltonians --- forms a critical ingredient of the
analytical framework presented below, allowing us to uncover the
complex structures of eigenvalues and their associated
eigenvectors of the corresponding dynamical matrices in the space
of system's parameters.} Moreover, we also consider bosonic
systems in which damping and amplification are not in balance,
which is a typical situation of $\mathcal{PT}$-symmetric systems.
In this unbalanced case, relying on the results presented in
Ref.~\cite{Chimczak2023}, we may introduce a specific interaction
picture in which the average damping or amplification dynamics is
projected out and the remaining dynamics exhibits the features
found in $\mathcal{PT}$-symmetric systems.

Under these conditions, we may write the quadratic Hamiltonian of
the considered two-mode bosonic system in the interaction picture
as follows:
\begin{eqnarray}  
 \hat{H}_{2} &=& \left[ \hbar \epsilon \hat{a}_{1}^{\dagger}\hat{a}_{2}
   + \hbar \kappa \hat{a}_{1}\hat{a}_{2} \right] + \textrm{H.c.},
\label{1}   
\end{eqnarray}
where $\hat{a}_{j} $ ($ \hat{a}_{j}^{\dagger} $) stands for the
annihilation (creation) operator of the $j$th mode, $ j=1,2 $, $
\epsilon$ is the linear coupling strength between the modes, and
$\kappa$ is the nonlinear coupling strength between the modes.
Symbol H.c.~replaces the Hermitian-conjugated terms. Whereas the
linear coupling originates in the spatial overlap of the mode
electric-field amplitudes, the nonlinear coupling arises in the
three-mode parametric process with strong pumping \cite{Boyd2003}.
The damping (amplification) of modes occurs as a consequence of
the interaction with the reservoir whose two-level atoms are in
the ground (excited) state \cite{PerinaJr2022a}. Projecting out
the reservoir two-level atoms, we are left with the damping
(amplification) rate $ \gamma_j $ of $ j $th damped (amplified)
mode and the corresponding Langevin stochastic operator forces, $
\hat{L}_j $ and $ \hat{L}_j^\dagger $, in the dynamical
Heisenberg-Langevin equations. We note that properties of the
Langevin operator forces differ for the damping and amplification
processes and they have to be chosen such that the field-operator
commutation relations are fulfilled. This results in the
corresponding fluctuation-dissipation theorems
\cite{Vogel2006,Perina1991} formulated within the
Heisenberg-Langevin formalism.

Using the two-mode Hamiltonian in Eq.~(\ref{1}), we derive the
Heisenberg-Langevin equations written for the vector $
\hat{\bm{a}}=\left[\hat{\bm{a}}_{\bm{1}}, \hat{\bm{a}}_{\bm{2}}
\right]^T \equiv \left[\hat{a}_{1},
\hat{a}_{1}^{\dagger},\hat{a}_{2},\hat{a}_{2}^{\dagger} \right]^T$
of field operators and vector $\hat{\bm{L}}=\left[\hat{L}_{1},
\hat{L}_{1}^{\dagger},\hat{L}_{2},\hat{L}_{2}^{\dagger} \right]^T$
of the Langevin operator forces as follows \cite{PerinaJr2000}:
\begin{eqnarray}  
 \frac{d\hat{\bm{a}}}{dt} & = & -i \bm{M^{\left(2 \right)}} \hat{\bm{a}} +\hat{\bm{L}}
\label{2}.   
\end{eqnarray}
In Eq.~(\ref{2}), the dynamical matrix $  \bm{M^{(2)}} $,
\begin{eqnarray}   
 \bm{M^{(2)}}  & = & \left[
 \begin{array}{cc}
  -i \bm{\tilde{\gamma}_1} & \bm{\xi} \\
  \bm{\xi} & -i \bm{\tilde{\gamma}_2}
 \end{array} ,
 \right] \label{3}  
\end{eqnarray}
is expressed in terms of $ 2\times 2 $ submatrices $ \bm{\tilde{
\gamma}_j} $, $ j=1,2 $, and $ \bm{\xi} $ is defined as:
\begin{eqnarray}   
 \bm{\tilde{\gamma}_j}=\left[
  \begin{array}{cc}
   \gamma_j/2 & 0 \\
    0 & \gamma_j/2  \\
  \end{array} \right], \hspace{5mm}
 \bm{\xi}=\left[
  \begin{array}{cc}
   \epsilon & \kappa \\
   -\kappa & -\epsilon \\
  \end{array} \right],
  \label{4}    
\end{eqnarray}
where $\gamma_j$ is the damping or amplification rate of the mode
$j$ that is accompanied by the corresponding Langevin operator
forces.

The $ 2\times 2 $ matrices, $ \bm{\tilde{ \gamma}_j} $ ($ j=1,2 $)
and $ \bm{\xi} $, given in Eq.~(\ref{4}) can be simultaneously
diagonalized using the diagonalization transformation appropriate
to the matrix $ \bm{\xi} $ as the remaining two matrices are
linearly proportional to the unity matrix and, thus, are not
modified by the transformation. This diagonalization
transformation then decomposes the $ 4\times 4 $ matrix $
\bm{M^{(2)}} $ in Eq.~(\ref{3}) into the direct sum of two
independent $ 2\times 2 $ matrices belonging to the eigenvalues
\Blue{$ \lambda^{\xi}_{1} $} and \Green{$ \lambda^{\xi}_{2} $} of
the matrix $ \bm{\xi} $ [see Eq.~(\ref{7}) below] and having the
structure of the original {$ 2\times 2 $} matrix $ \bm{M^{(2)}} $:
\begin{eqnarray}  
 \bm{M^{(2)} }
=  \left[
  \begin{array}{cccccccc}
   \Blue{-i {\gamma}_1/2} & 0    & \Blue{\lambda^{\xi}_{1}} & 0   \\
   0 & \Green{ -i {\gamma}_1/2} &  0 & \Green{\lambda^{\xi}_{2} }   \\
   \Blue{\lambda^{\xi}_{1}} & 0 & \Blue{-i {\gamma}_2/2} & 0 \\
   0 & \Green{ \lambda^{\xi}_{2}} &  0 & \Green{ -i {\gamma}_2/2}
  \end{array}
 \right] .
\nonumber   
\end{eqnarray}
This step is critical to our analytical approach, and it is
applied analogously to systems with a larger number of bosonic
modes. It enables an effective reduction in the dimensionality of
the diagonalization problem, thereby allowing the derivation of
comprehensive analytical expressions essential for analyzing QEPs
and QDPs. Denoting a general eigenvalue of the $ 2\times 2 $
matrix $ \bm{\xi} $ as $ \xi $, we may express the
eigenfrequencies $ \lambda_{1,2}^{M^{(2)}} $ and eigenvectors $
\bm{y_{1,2}^{M^{(2)}} } $ of the decomposing $ 2\times 2 $
matrices of the $ 4\times 4 $ matrix $\bm{M^{(2)}}$ in
Eq.~(\ref{2}) in the following common form:
\begin{eqnarray}  
 {\lambda_{1,2}^{M^{(2)}} }& =&-i \gamma_{+} \mp \beta
\label{5}   
\end{eqnarray}
and
\begin{eqnarray}   
 \bm{y_{1,2}^{M^{(2)}} }& =&\left[-\frac{i
 \gamma_{-}\pm \beta}{{\xi}} , 1 \right]^T,
\label{6}   
\end{eqnarray}
where $ 4\gamma_{\pm} = \gamma_{1} \pm \gamma_{2}$ and $
\beta^2={{\xi}^2-\gamma_{-}^2}$.

The eigenvalues $ \lambda^{\xi}_{1,2} $ and the corresponding
eigenvectors $ \bm{y^{\xi}_{1,2}} $ of the matrix $\bm{\xi}$ are
simply derived in the form:
\begin{eqnarray} 
 \lambda^{\xi}_{1,2}& =&\mp \zeta
\label{7}  
\end{eqnarray}
and
\begin{eqnarray}  
 \bm{y^{\xi}_{1,2}}& =&\left[-\frac{\epsilon \mp \zeta}{\kappa} , 1 \right]^T,
\label{8} 
\end{eqnarray}
where $\zeta=\sqrt{\epsilon^2-\kappa^2}$.

Combing the above two results for the matrix diagonalization, we
can express the eigenvalues $\Lambda^{M^{(2)}} $ of the $ 4\times
4 $ matrix $\bm{M^{(2)}}$ in Eq.~(\ref{2}) as
$\Lambda_{1,3}^{M^{(2)}} = {\lambda_{1,2}^{M^{(2)}} } $ for $
{\xi}=\lambda^{\xi}_{1} $ and $\Lambda_{2,4}^{M^{(2)}}
={\lambda_{1,2}^{M^{(2)}} } $ for ${\xi}=\lambda^{\xi}_{2} $,
i.e.:
\begin{eqnarray}   
 \Lambda_{1}^{M^{(2)}} & = & \Lambda_{2}^{M^{(2)}}=-i \gamma_{+}-\beta, \nonumber \\
 \Lambda_{3}^{M^{(2)}}& = &\Lambda_{4}^{M^{(2)}}=-i \gamma_{+}+\beta,
\label{9}   
\end{eqnarray}
and $\beta= \sqrt{\zeta^2- \gamma_{-}^2} $.  {We note that, in
general, we use capital Greek letters $\Lambda$ to denote the
eigenvalues of the dynamical matrices ${\bm M}$ in their full
dimensions, and lowercase Greek letters $\lambda$ for the
eigenvalues in the reduced (half) dimensions.} We also note that
the average damping or amplification rate $ \gamma_+ = 0 $ in the
usual $\mathcal{PT}$-symmetric systems,  {that are, however,
only specific cases in our general calculations.}

\begin{figure}  
 \vspace{-20mm}
 \begin{centering}
  \includegraphics[width=0.95\hsize]{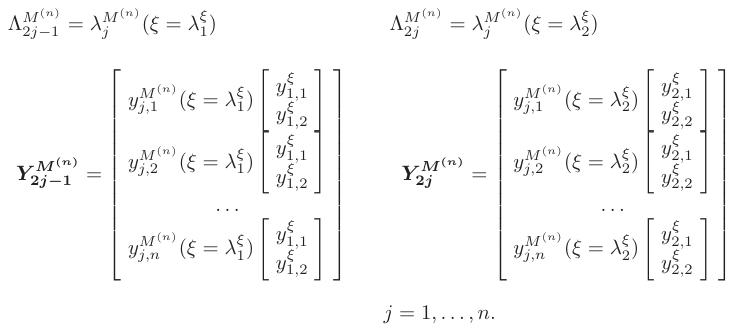}
 \end{centering}
 \vspace{-130mm}
 \caption{ {Schematic diagram of the structure of eigenvalues $
 \Lambda_{j}^{M^{(n)}}$ and the corresponding eigenvectors
  $ \bm{Y_{j}^{M^{(n)}}} $, $ j=1,\ldots,2n $, of a general dynamical $ 2n\times  2n $ matrix
  $ \bm{M^{(n)}} $ built from the eigenvalues $
   \lambda_{k}^{M^{(n)}}$ and the corresponding eigenvectors
  $ \bm{y_{k}^{M^{(n)}}} $, $ k=1,\ldots,n $, of the matrix
  $ \bm{M^{(n)}} $ considered as an $ n\times  n $ matrix composed of
  $ 2\times  2 $ submatrices at the positions of its elements; $
  n=2,3,\ldots $. The eigenvalues $ \lambda^{\xi}_{1,2} $ and the corresponding
  eigenvectors $ \bm{y^{\xi}_{1,2}} $ belong to the submatrix $
  \bm{\xi} $.}}
\label{fig2}
\end{figure}

Similarly as the eigenvalues, we obtain the eigenvectors along the
formulas
\begin{eqnarray}  
 \bm{Y_{j}^{M^{(2)}}} &=& \left[
  {y_{1,1}^{M^{(2)}} }({\xi}=\lambda^{\xi}_{j}) \bm{y^{\xi}_{j}} ,
  {  y_{1,2}^{M^{(2)}} }({\xi}=\lambda^{\xi}_{j}) \bm{y^{\xi}_{j}}
  \right]^T , \nonumber \\
  & &  \hspace{15mm} {\rm for}\;\; j = 1,2, \nonumber \\
 \bm{Y_{j}^{M^{(2)}}} &=& \left[ {y_{2,1}^{M^{(2)}}
  }({\xi}=\lambda^{\xi}_{j-2} ) \bm{y^{\xi}_{j-2}} ,
  {y_{2,2}^{M^{(2)}} }({\xi}=\lambda^{\xi}_{j-2}) \bm{y^{\xi}_{j-2}}
  \right]^T, \nonumber \\
  & &  \hspace{15mm} {\rm for}\;\; j = 3,4,
\label{10}
\end{eqnarray}
in the form:
\begin{eqnarray} 
 \bm{Y_{1,2}^{M^{(2)}}}&= &
  \left[ \begin{array}{cccccc}
  \frac{(\mp\epsilon + \zeta) \chi}{\kappa\zeta}, & \pm\frac{\chi}{\zeta}, & -\frac{(\epsilon \mp \zeta)}{\kappa}, & 1
   \end{array}\right]^T , \nonumber \\
 \bm{Y_{3,4}^{M^{(2)}}} &= &
  \left[ \begin{array}{cccccc}
  \frac{(\pm\epsilon - \zeta) \chi^{*}}{\kappa\zeta}, & \mp\frac{\chi^{*}}{\zeta}, & -\frac{(\epsilon \mp \zeta)}{\kappa}, & 1
   \end{array}\right]^T ,
\label{11}   
\end{eqnarray}
where $\chi=i \gamma_{-}+\beta $.  {The structure of
eigenvalues and eigenvectors of a general dynamical $ 2n\times  2n
$ matrix $ \bm{M^{(n)}} $ formed by the eigenvalues $
\lambda_{k}^{M^{(n)}}$ and eigenvectors $ \bm{y_{k}^{M^{(n)}}} $
of the $ n\times n $ matrix $ \bm{M^{(n)}} $ expressed using $
2\times 2 $ submatrices and the eigenvalues $ \lambda^{\xi}_{1,2}
$ and eigenvectors of the submatrix $ \bm{\xi} $ is shown in
Fig.~\ref{fig2}.}

The formula in Eq.~(\ref{9}) provides the two pairs of coinciding
eigenvalues. However, Eq.~(\ref{11}) for the corresponding
eigenvectors reveal no eigenvector degeneracy in general. On the
other hand, $ \chi $ is purely imaginary for ${\beta}=0$, which
leads to $\bm{Y_{1}^{M^{(2)}}}=\bm{Y_{3}^{M^{(2)}}}$ and
$\bm{Y_{2}^{M^{(2)}}}=\bm{Y_{4}^{M^{(2)}}}$, while having all the
eigenvalues the same. So we observe the second-order degeneration
in the Hilbert space that is formed by two second-order QEPs with
identical eigenvalues. We, thus, have a QHP with second-order DD
and ED in this case. The condition ${\beta}=0$ implies that the
eigenvalues and eigenvectors of the $ 2\times 2 $ matrix $
\bm{M^{(2)}} $ given in Eqs.~(\ref{5}) and (\ref{6}) coincide. The
second-order ED, thus, originates in the $ 2\times 2 $ form of
matrix  $ \bm{M^{(2)}} $. We note that this degeneracy can be
verified by transforming the matrix $ \bm{M^{(2)}} $ for $\beta=0$
into its Jordan form
\begin{equation}   
  \bm{J_{M^{(2)}}} = \left[
  \begin{array}{cc}
   -i\gamma_{+} & 1 \\
    0 & -i\gamma_{+}  \\
  \end{array} \right].
\end{equation}
On the other hand, the eigenvalues of $ 2\times 2 $ submatrix $
\bm{\xi} $ written in Eq.~(\ref{7}) point out at the origin of DD:
They differ just by the sign, but they lead to the same value of $
\beta$, i.e., to the same eigenvalue $ \lambda_{1,2}^{M^{(2)}} $
of the $ 2\times 2 $ matrix $ \bm{M^{(2)}} $. DD is then implied
by the fact that the eigenvectors $ \bm{y^{\xi}_{1,2}} $ of  $
2\times 2 $ submatrix $ \bm{\xi} $ in Eq.~(\ref{8}) differ for $
\zeta \neq 0 $.

These findings about the structure of the matrices describing the
dynamical equations have their counterpart in the structure of the
analyzed two-mode $\mathcal{PT}$-symmetric system. The  $ 2\times
2 $ matrix $ \bm{\xi} $, defined in Eq.~(\ref{4}), connects pairs
of the annihilation and creation operators of different modes. As
such it forms the basic building block of more complex
$\mathcal{PT}$-symmetric systems, together with the damping and/or
amplification matrices $ \bm{\tilde{\gamma}_j} $ in Eq.~(\ref{4}).
In more complex $\mathcal{PT}$-symmetric systems, the dependencies
of the eigenvalues of the dynamical  {$ n\times n $} matrices $
\bm{ M^{(n)}} $ ($ n=2,3,\ldots $) on the eigenvalues $
\lambda^{\xi}_{1,2} $ of the $ \bm{\xi} $ matrix are typically
quadratic. This, thus, results in the observation of second-order
DD in the dynamical features of the matrices $ \bm{ M^{(n)}} $. In
this case, the eigenvalue analysis of the considered systems
considerably simplifies and we may restrict our attention to only
the $ n\times n $ matrices $ \bm{ M^{(n)}} $  {with their
elements in the form of $ 2\times 2 $ submatrices}, when a
detailed eigenvalue analysis is performed.  {The consequences
of the eigenvalue analysis are then combined with the above
second-order DD.}

\begin{figure*}  
 \centerline{  \includegraphics[width=0.3\hsize]{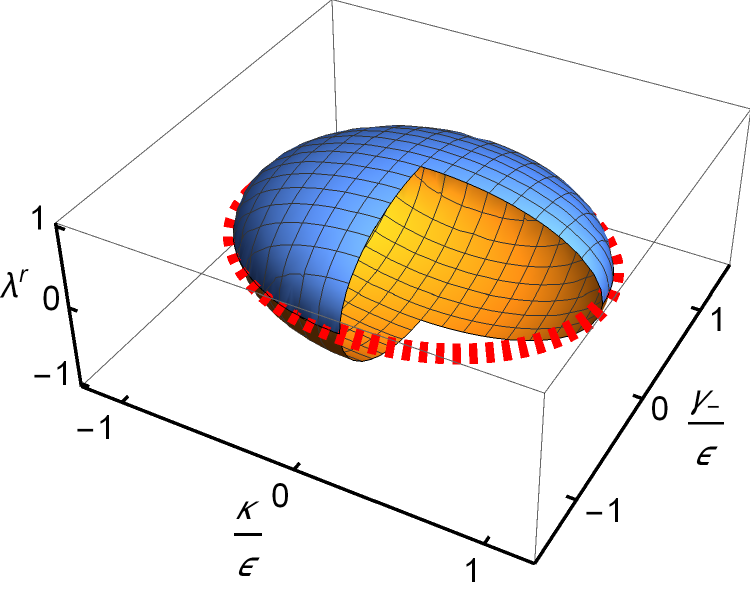}
  \hspace{3mm}  \includegraphics[width=0.3\hsize]{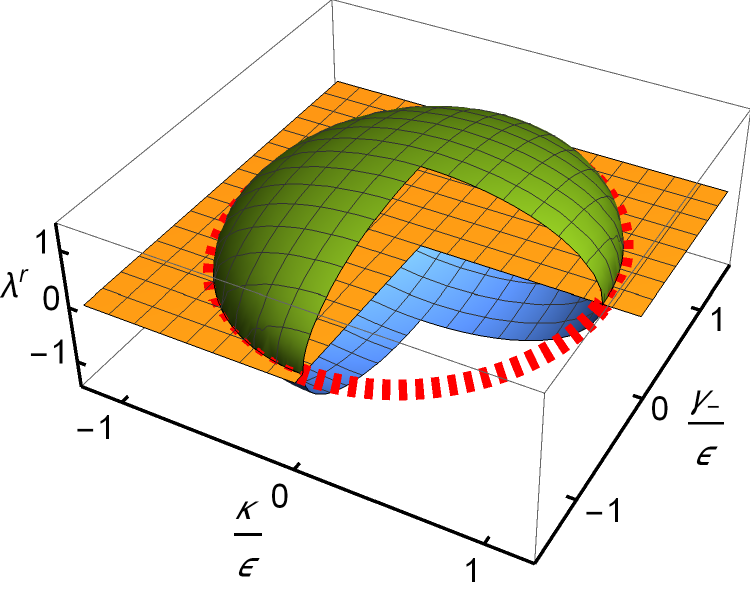}
  \hspace{3mm} \includegraphics[width=0.3\hsize]{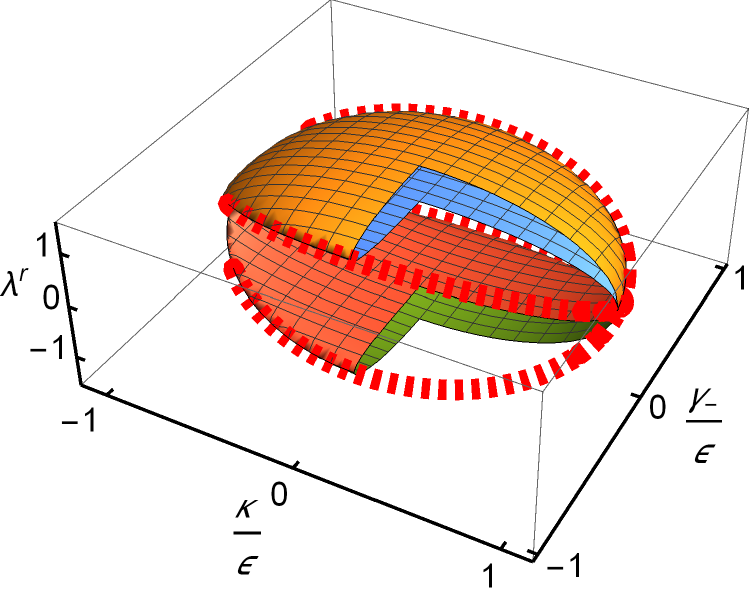}  }
 \centerline{ {\small (a)} \hspace{0.3\hsize}  {\small (b)} \hspace{0.3\hsize} {\small (c)} \hfill }
 \vspace{4mm}

 \centerline{  \includegraphics[width=0.3\hsize]{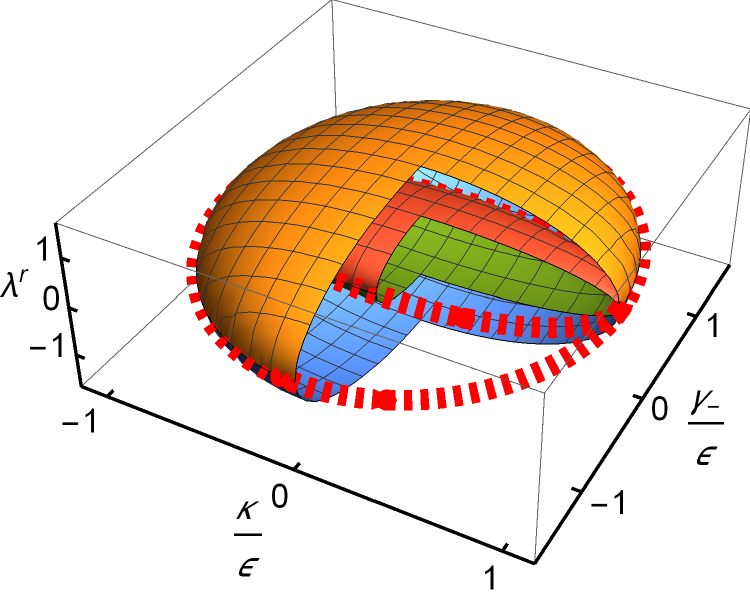}
  \hspace{3mm}  \includegraphics[width=0.3\hsize]{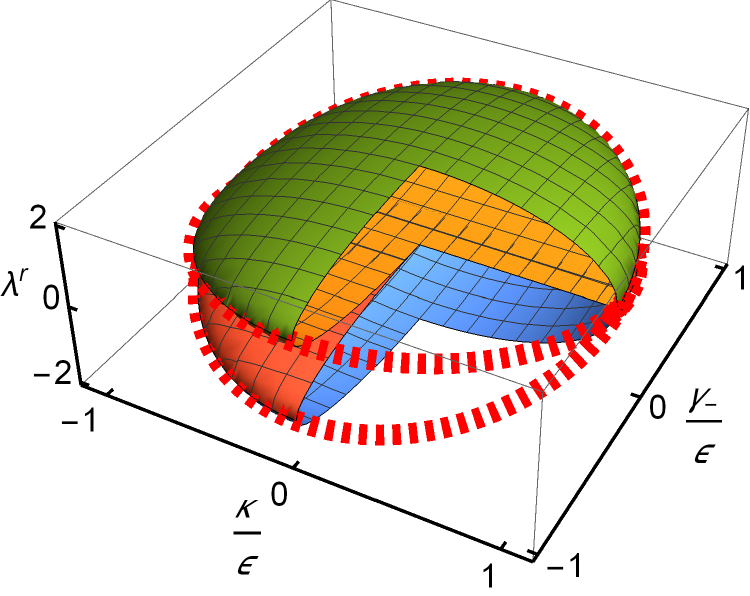}
  \hspace{3mm} \includegraphics[width=0.3\hsize]{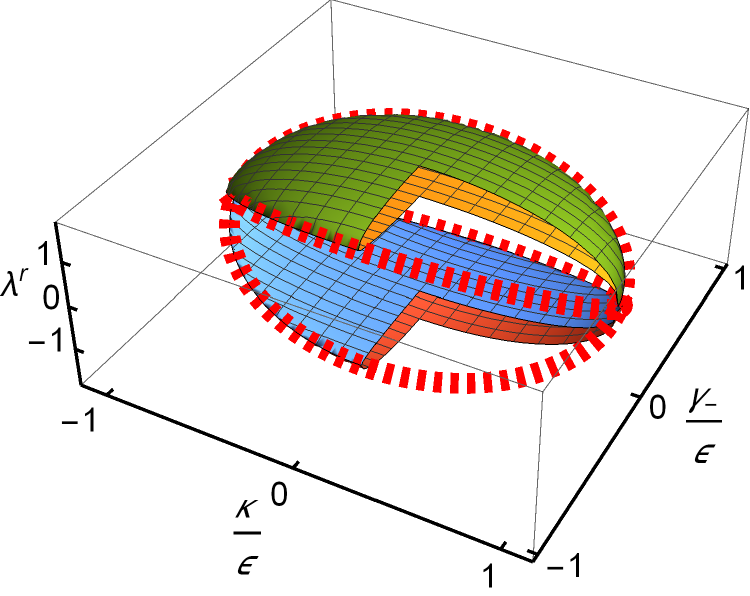}  }
 \centerline{ {\small (d)} \hspace{0.3\hsize}  {\small (e)} \hspace{0.3\hsize} {\small (f)} \hfill }
 \vspace{4mm}

 \centerline{  \includegraphics[width=0.3\hsize]{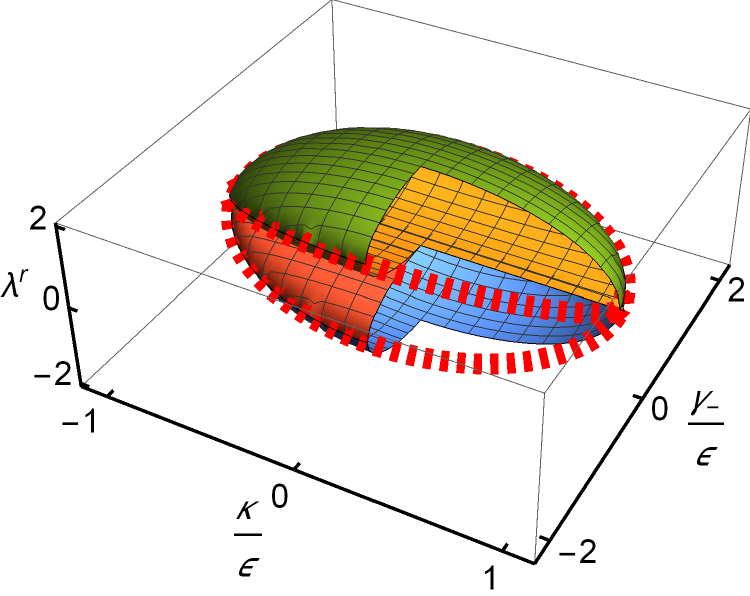}
  \hspace{3mm}  \includegraphics[width=0.3\hsize]{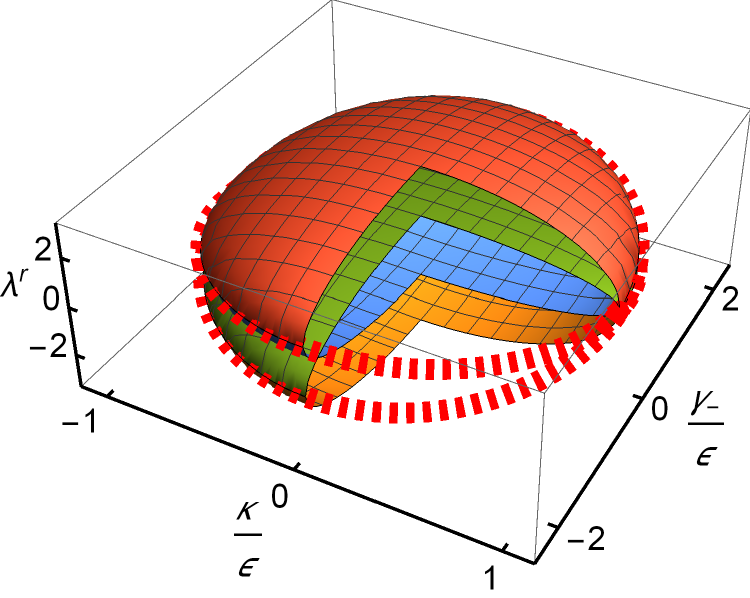}}
 \centerline{ \hspace{0.15\hsize} {\small (g)} \hspace{0.3\hsize}  {\small (h)} \hfill }

 \caption{Real parts $ \lambda^{\rm r} $ of the eigenvalues (a) $ \lambda_{1,2}^{M^{(2)}} $
  of the matrix $ \bm{M^{(2)}} $, given in Eq.~(\ref{5}), for the two-mode bosonic system,
  (b) $ \lambda_{1,2,3}^{M^{(3)}} $ of the matrix $ \bm{M^{(3)}} $, given in Eq.~(\ref{23}),
  for the three-mode linear bosonic system,
  (c) [(d)] $ \lambda_{1,\ldots,4}^{M^{(4)}_{\rm l1}} $ [$ \lambda_{1,\ldots,4}^{M^{(4)}_{\rm l2}} $]
   of the matrix $ \bm{M^{(4)}_{\rm l1}} $ [$ \bm{M^{(4)}_{\rm l2}}
   $], given in Eq.~(\ref{30}) [(\ref{34})] for the four-mode linear bosonic system
   with neighbor modes having equal [different] damping and/or amplification
   rates,
  (e) $ \lambda_{1,\ldots,4}^{M^{(4)}_{\rm c1}} $ of the matrix
  $ \bm{M^{(4)}_{\rm c1}} $, given in Eq.~(\ref{40}), for the four-mode circular bosonic system
  with neighbor modes having equal damping and/or amplification
  rates,
  (f) $ \lambda_{2,\ldots, 5}^{M^{(5)}_{\rm l}} $ of the matrix
  $ \bm{M^{(5)}_{\rm l}} $, given by Eq.~(\ref{47}), for the five-mode linear bosonic system,
  and (g,h) $ \lambda_{2,\ldots, 5}^{M^{(5)}_{\rm p}} $ of the matrix
  $ \bm{M^{(5)}_{\rm p}} $, given in Eq.~(\ref{54}), for the five-mode pyramid bosonic
  system assuming (g) $\beta_1=0$  and (h) $\beta_2=0$.
  The eigenvalues are drawn in the parameter space
  $ (\kappa/\epsilon,\gamma_-/\epsilon) $, where $ \epsilon $ ($ \kappa
  $) is the linear (nonlinear) coupling strength and $ \gamma_- $ the difference
  of the damping/amplification rates in individual models. Dashed red curves indicate
  the positions of the QHPs given by (a,e,g) Eq.~(\ref{13}), (b)
  Eq.~(\ref{25}), (c) Eq.~(\ref{32}), (d) Eq.~(\ref{36}), (f) Eq.~(\ref{49}), and
  (h) Eq.~(\ref{56}).}
\label{fig3}
\end{figure*}

The condition ${\beta}=0$ for observing QHPs can be analyzed in
the space of system parameters $
(\epsilon,\kappa,\gamma_1,\gamma_2) $ as follows. The quadratic
Hamiltonian $ \hat{H}^{(2)} $ in Eq.~(\ref{1}) provides the linear
Heisenberg-Langevin equations that allow for the temporal
rescaling $ \epsilon t $, i.e., only the relative parameters $
(\kappa/\epsilon,\gamma_1/\epsilon,\gamma_2/\epsilon) $ suffice in
characterizing the system dynamics. Moreover, the structure of the
analyzed systems is such that the average damping or amplification
rate $ \gamma_+ $ influences equally only the eigenvalues, but it
does not modify the eigenvectors. This makes the space of the
system parameters effectively two-dimensional with the spanning
parameters $ (\kappa/\epsilon,\gamma_-/\epsilon) $. Using these
parameters, the condition $ \beta=0 $ is expressed as follows:
\begin{equation}   
 \frac{\kappa^2}{\epsilon^2}+ \frac{\gamma_{-}^2}{\epsilon^2}=1.
\label{13}
\end{equation}
Thus, the QHPs form a circle with the unit radius in the space $
(\kappa/\epsilon,\gamma_-/\epsilon) $, as shown by  {the red
dashed curve} in Fig.~\ref{fig3}(a). We note that, at this circle,
there is a specific point at $\gamma_{-}=0$ in which all four
eigenvectors, given in Eq.~(\ref{10}), are the same which gives
raise to the fourth-order QEP. However, this corresponds to the
system in which both modes are equally damped or amplified. It is
worth noting that the eigenvectors of both matrices $ \bm{M^{(2)}}
$, given in Eq.~(\ref{3}), and $ \bm{\xi} $, in Eq.~(\ref{4}), are
degenerated at this point.

Deeper insight into the structure of higher-order FOMs, as well as
the system dynamics, can be obtained once we transform the
Heisenberg-Langevin equations (\ref{2}) into the form in which the
dynamical matrix has the diagonal form. Using the transformation
matrix $\bm{P}=\left[\bm{Y_{1}^{M^{(2)}}},
\bm{Y_{2}^{M^{(2)}}},\bm{Y_{3}^{M^{(2)}}},\bm{Y_{4}^{M^{(2)}}}
\right]$ formed from the eigenvectors in Eq.~(\ref{10}), we define
the corresponding field operators $\hat{\bm{b}}=\left[\hat{b}_{1},
\hat{b}_{2},\hat{b}_{1}^{\dagger},\hat{b}_{2}^{\dagger}
\right]^T$, the Langevin operator forces
$\hat{\bm{K}}=\left[\hat{K}_{1},
\hat{K}_{2},\hat{K}_{1}^{\dagger},\hat{K}_{2}^{\dagger}
\right]^T$, and the diagonal dynamical matrix $ \bm{\Lambda^{(2)}}
$:
\begin{eqnarray}   
 \bm{\Lambda^{(2)}}= \bm{P}^{-1}\bm{M^{(2)}}
 \bm{P}, \hspace{3mm} \hat{\bm{b}}=\bm{P}^{-1}\hat{\bm{a}}, \hspace{3mm}
 \hat{\bm{K}}=\bm{P}^{-1}\hat{\bm{L}}.
\label{14}
\end{eqnarray}
We note that the order of elements in the operator vector $
\hat{\bm{b}} $ (and similarly in the vector $ \hat{\bm{K}} $ of
the accompanying Langevin operator forces) is given by the
numbering of the eigenvalues in Eq.~(\ref{9}) and the
corresponding diagonalization transform.

In the transformed basis, the Heisenberg-Langevin equations take
the form:
\begin{eqnarray}   
 \frac{d\hat{\bm{b}}}{dt} & = & -i \bm{\Lambda^{(2)}} \hat{\bm{b}}
 +\hat{\bm{K}}.
\label{15}    
\end{eqnarray}
We note that the positions of the newly-defined annihilation and
creation operators in the vector $ \hat{\bm{b}} $, as well as the
positions of the accompanying Langevin operator forces in the
vector $ \hat{\bm{K}} $, differ from those in the original vectors
$ \hat{\bm{a}} $ and $ \hat{\bm{L}} $ defined above Eq.~(\ref{2}).
The solution to Eq.~(\ref{15}) is expressed as:
\begin{eqnarray}   
 \hat{\bm{b}}(t) &=& \exp(-i \bm{\Lambda^{(2)}}t)
  \hat{\bm{b}}(0) \nonumber \\
   & & \mbox{} + \int_{0}^{t} dt' \exp[-i \bm{\Lambda^{(2)}}(t-t')] \hat{\bm{K}}(t').
\label{16}
\end{eqnarray}
Using the inverse transformation, we arrive at the solution for
the operators $ \hat{\bm{a}} $, which reads:
\begin{eqnarray}   
 \hat{\bm{a}}(t) &=&  \bm{P} \exp(-i \bm{\Lambda^{(2)}}t)
   \bm{P}^{-1} \hat{\bm{a}}(0) \nonumber \\
  & & \mbox{} +  \int_{0}^{t} dt'  \bm{P}\exp[-i \bm{\Lambda^{(2)}}(t-t')]  \bm{P}^{-1} \hat{\bm{L}}(t').
\label{17}
\end{eqnarray}

The solution in Eq.~(\ref{17}) of the Heisenberg-Langevin
equations in the diagonal form together with the assumption of the
Gaussian Markovian character of the Langevin operator forces
allows us to follow the dynamics of FOMs of an arbitrary order. We
note that, from the point of view of eigenfrequencies, the FOMs of
a given order form a closed dynamical system \cite{PerinaJr2022a}.
In this dynamics, we observe the genuine QEPs and QHPs that are
derived from the inherited QEPs and QHPs discussed above. In
general, the higher is the FOM order, the higher are the observed
EDs and DDs at QEPs and QHPs.  {These degeneracies were
systematically studied in Ref.~\cite{PerinaJr2022a} up to
fourth-order FOMs. They form intricate structures within FOM
spaces of defined order, where specific QEPs and QHPs emerge. Such
spaces and structures can be effectively utilized to simulate
various physical phenomena
\cite{Arkhipov2021,Arkhipov2023}. In
Appendix~\ref{AppB}, we systematically summarize these
degeneracies for up to second-order FOMs, providing a comparative
overview of the QEP and QHP degeneracies observed in the first-
and second-order FOM dynamics.}

We note that the quadratic form of Hamiltonian $ \hat{H}_2 $, in
Eq.~(\ref{1}), is not the most general one. It can be extended by
considering additional terms describing local squeezing in modes 1
and 2 described by a constant $ g $, as it was done in
Ref.~\cite{PerinaJr2022a}:
\begin{eqnarray} 
 \hat{\tilde{H}}_2 = \left[ \hbar \epsilon \hat{a}^\dagger_1 \hat{a}_2 +
  \hbar \kappa \hat{a}_1\hat{a}_2 + \sum_{j=1,2} \hbar g\hat{a}_j^{\dagger
  2}/2\right]  + {\rm H.c.}
\label{18}
\end{eqnarray}
However, the structure of the dynamical matrix $ \bm{M^{(2)}} $,
built from the $ 2\times 2 $ submatrices, is broken for nonzero
values of the constant $ g $. This results in the loss of the
second-order DD that follows from the structure of the $ \bm{\xi}
$ and $ \bm{\tilde{\gamma}_j} $ matrices written in Eq.~(\ref{4}).
As a consequence, only second-order inherited QEPs occur for $
g\neq 0 $ in the parameter space $
(\kappa/\epsilon,\gamma_-/\epsilon,g/\epsilon) $. Detailed
analysis of the positions of inherited QEPs, as well as EDs and
DDs of QEPs and QHPs observed in the FOMs dynamics was provided in
Ref.~\cite{PerinaJr2022a}. Similar reduction of eigenvalue and
eigenvector degeneracies in the Hilbert space after including
these terms was observed in more complex bosonic systems. We
further pay attention only to the systems described by quadratic
Hamiltonians without these terms.

\section{Three-mode bosonic system}
\label{sec3}

The above-analyzed two-mode system provided QHPs with second-order
ED and DD. Whereas the DD originates in the mutual coupling
between the two modes (described by the strengths $ \epsilon $ and
$ \kappa $), the ED arises from different strengths of damping and
amplification of the modes and a given system configuration.
Bosonic systems composed of more than two modes give a promise for
revealing QHPs with a larger number of EDs and DDs  {and
various configurations of the coexistence of several QHPs with
different EDs and DDs.} However, the question is how to choose a
suitable configuration of interactions among the modes and their
damping and/or amplification rates. Whereas the parity symmetry is
useful in seeking promising geometries, the temporal symmetry
allows to define suitable relations among the damping and
amplification rates. In this section, we begin our investigation
by considering three-mode bosonic systems in the linear
configuration, and define specific conditions under which QHPs
occur. In the following sections, we extend our analysis to the
four- and five-mode systems. The geometries together with their
characteristic parameters are schematically shown in
Fig.~\ref{fig1}. We also identify the positions of QHPs in the
corresponding parameter spaces. Moreover, we analyze the
occurrence of genuine QEPs and QHPs and determine their EDs and
DDs from the point of view of the dynamics of second-order FOMs.

Among the bosonic systems with three modes we succeeded in
identifying QHPs in the linear configuration that is shown in
Fig.~\ref{fig1}(b).  {We note that a linear non-Hermitian
Hamiltonian (i.e. semiclassical) system in a similar configuration
was experimentally realized in Ref.~\cite{Hodaei2017} and used for
higher-order exceptional-point enhanced sensing.} We note that a
QEP with third-order ED was identified in Ref.~\cite{Jing2017} in
an optomechanical system of three interacting bosonic modes.
Hamiltonian $ \hat{H}_{3} $ of such a system can be written as
follows:
\begin{eqnarray}    
 \hat{H}_{3} &=& \left[ \hbar \epsilon \hat{a}_{1}^{\dagger}\hat{a}_{2}
  + \hbar \epsilon \hat{a}_{2}^{\dagger}\hat{a}_{3}  + \hbar \kappa \hat{a}_{1}\hat{a}_{2}+
  \hbar \kappa \hat{a}_{2}\hat{a}_{3} \right] + \textrm{H.c.}\nonumber \\
\label{19}   
\end{eqnarray}

Using the $ 2\times 2 $ submatrices $ \bm{\tilde{\gamma}_j} $, $
j=1,2,3 $, describing mode damping or amplification, and $
\bm{\xi} $ defined in Eq.~(\ref{4}), we derive the
Heisenberg-Langevin equations in the form
\begin{eqnarray}   
 \frac{d\hat{\bm{a}}}{dt} & = & -i \bm{M^{(3)}} \hat{\bm{a}} +\hat{\bm{L}},
\label{20}   
\end{eqnarray}
where
\begin{eqnarray}   
 \bm{M^{(3)}}  & = & \left[
  \begin{array}{ccc}
   -i \bm{\tilde{\gamma}_1} & \bm{\xi} & \bm{0} \\
   \bm{\xi} & -i \bm{\tilde{\gamma}_2} & \bm{\xi}\\
  \bm{0}  & \bm{\xi} & -i \bm{\tilde{\gamma}_3}\\
  \end{array}
  \right] ,
\label{21} 
\end{eqnarray}
and $\bm{0} $  denotes the $ 2\times 2 $ null matrix, while
$\hat{\bm{a}}=\left[\hat{\bm{a}}_{\bm{1}}, \hat{\bm{a}}_{\bm{2}},
\hat{\bm{a}}_{\bm{3}}  \right]^T \equiv\left[\hat{a}_{1},
\hat{a}_{1}^{\dagger},\hat{a}_{2},\hat{a}_{2}^{\dagger},\hat{a}_{3},\hat{a}_{3}^{\dagger}
\right]^T$ and $\hat{\bm{L}}=\left[\hat{L}_{1},
\hat{L}_{1}^{\dagger},\hat{L}_{2},\hat{L}_{2}^{\dagger},\hat{L}_{3},\hat{L}_{3}^{\dagger}
\right]^T$.

 {The condition
\begin{equation}   
  2\gamma_2 =\gamma_1+\gamma_3
\label{22}
\end{equation}
is required to have the eigenvalues $ \lambda_{j}^{M^{(3)}} $ ($
j=1,2,3 $) of the dynamical $ 3\times 3 $ matrix $\bm{M^{(3)}}$ in
Eq.~(\ref{21}) with a common damping or amplification rate $
\gamma_+ $. This is necessary for observing possible spectral
exceptional and diabolical degeneracies.}

 {We note that condition in Eq.~(\ref{22}) also encompasses
passive ($ \gamma_j > 0 $ for $ j=1,2,3 $) and active $ \gamma_j <
0 $ for $ j=1,2,3 $ $\mathcal{PT}$-symmetric systems.} Under this
condition we may write
\begin{eqnarray}  
  \lambda_{1}^{M^{(3)}} & =& -i \gamma_+ ,\quad
  \lambda_{2,3}^{M^{(3)}} =-i \gamma_+ \mp \beta,
\label{23}   
\end{eqnarray}
using $ 4\gamma_{\pm} = \gamma_{1} \pm \gamma_{3}$ and $
\beta^2=2\xi^2-\gamma_{-}^2 $; $ \xi $ being an eigenvalue of the
matrix $ \bm{\xi} $ in Eq.~(\ref{4}). The eigenvectors
corresponding to the eigenvalues $ \lambda_{j}^{M^{(3)}} $ in
Eq.~(\ref{23}) are derived in the form:
\begin{eqnarray}  
 \bm{y_{1}^{M^{(3)}} } & =&\left[-1, -\frac{i \gamma_{-} }{ \xi } , 1 \right]^T, \nonumber \\
 \bm{y_{2,3}^{M^{(3)}} }& =&\left[1 +\frac{i \gamma_{-} (i
 \gamma_{-} \pm\beta) }{ \xi^2 } ,-\frac{i \gamma_{-} \pm
 \beta }{\xi} , 1 \right]^T.
\label{24}   
\end{eqnarray}

Provided that $\beta=0$, the three eigenvalues $
\lambda_{j}^{M^{(3)}} $ for $ j=1,2,3 $ in Eq.~(\ref{23}), as well
as the corresponding eigenvectors in Eq.~(\ref{24}), are equal
and, thus, third-order QEPs of the  {$ 3\times 3 $} matrix $
\bm{M^{(3)}} $ are found. Taking into account the structure of the
submatrices of the   {$ 6\times 6 $} matrix $ \bm{M^{(3)}} $,
these QEPs are in fact QHPs with second-order DD. Substituting for
the eigenvalues $ \xi $ of matrix $ \bm{\xi} $ from Eq.~(\ref{7}),
the condition $ \beta = \sqrt{ 2\zeta^2-\gamma_{-}^2 } = 0 $ for
having a QHP attains the form:
\begin{equation}   
 \frac{\kappa^2}{\epsilon^2}+ \frac{\gamma_{-}^2}{2\epsilon^2}=1.
\label{25}
\end{equation}
This condition is visualized by the red dashed curve in
Fig.~\ref{fig3}(b) that shows the real parts of the eigenvalues $
\lambda_{j}^{M^{(3)}} $ as they vary in the parameter space $
(\kappa/\epsilon,\gamma_-/\epsilon) $.

The analysis of QEPs and QHPs appropriate for the FOM dynamics
requires the construction of eigenvalues and eigenvectors of the
matrix $ \bm{ M^{(3)}} $ in its full $ 6\times 6 $ space combining
the eigenvalues and eigenvectors in Eqs.~(\ref{23}) and (\ref{24})
with those of the matrix $ \bm{\xi} $ given in Eqs.~(\ref{7}) and
(\ref{8}) using the general scheme shown in Fig.~\ref{fig2}. The
eigenvalues and eigenvectors are written in Appendix~\ref{AppA}.
The obtained eigenvectors then allow us to introduce new field
operators $\hat{\bm{b}}=\left[\hat{b}_{1},\hat{b}_{1}^{\dagger},
\hat{b}_{2},\hat{b}_{3},\hat{b}_{2}^{\dagger},\hat{b}_{3}^{\dagger}
\right]^T$ in which the dynamical  {$ 6\times 6 $} matrix
$\bm{M^{(3)}}$ of the Heisenberg-Langevin equations attains its
diagonal form $\bm{\Lambda^{(3)}}$, similarly as it was done in
Sec.~\ref{sec2} in the case of the two-mode system. This then
allows to reveal the spectral degeneracies in higher-order FOMs.
The QEPs and QHPs predicted in the dynamics of the first- and
second-order FOMs are summarized in Appendix~\ref{AppB} using
Tab.~\ref{tab2}.

\section{Four-mode bosonic systems}

Let us move to the investigations of four-mode bosonic systems, in
which we revealed QEPs and QHPs in two arrangements: linear and
circular. In the linear arrangement, we consider two
configurations that differ by the damping and/or amplification
rates assigned to the modes: Either the neighbor modes share their
damping and/or amplification rates or their rates differ. On the
other hand, equal damping and/or amplification rates of the
neighbor modes are needed in the circular arrangement to reveal
QEPs and QHPs.

\subsection{Linear configurations}

In the linear configuration, the quadratic Hamiltonian $
\hat{H}_{4,{\rm l}} $ of four-mode system is expressed in the form
\begin{eqnarray} 
 \hat{H}_{4,{\rm l}} &=& \Bigl[ \hbar \epsilon \hat{a}_{1}^{\dagger}\hat{a}_{2}
  + \hbar \epsilon \hat{a}_{2}^{\dagger}\hat{a}_{3}  + \hbar \epsilon \hat{a}_{3}^{\dagger}\hat{a}_{4}  + \hbar \kappa \hat{a}_{1}\hat{a}_{2}+ \hbar \kappa \hat{a}_{2}\hat{a}_{3} + \hbar \kappa \hat{a}_{3}\hat{a}_{4} \Bigr] + \textrm{H.c.}
\label{26}   
\end{eqnarray}
The corresponding Heisenberg-Langevin equations are derived as
follows:
\begin{eqnarray}   
 \frac{d\hat{\bm{a}}}{dt} & = & -i \bm{M^{(4)}_{\rm l}} \hat{\bm{a}} +\hat{\bm{L}},
\label{27}
\end{eqnarray}
Using the $ 2\times 2 $ submatrices defined in Eq.~(\ref{4}) we
write the dynamical  {$ 4\times 4 $} matrix $ \bm{M^{(4)}_{\rm
l}} $:
\begin{eqnarray}  
 \bm{M^{(4)}_{\rm l}}  & = & \left[
  \begin{array}{cccc}
   -i \bm{\tilde{\gamma}_1} & \bm{\xi} & \bm{0} & \bm{0}\\
   \bm{\xi} & -i \bm{\tilde{\gamma}_2} & \bm{\xi}& \bm{0}\\
   \bm{0}  & \bm{\xi} & -i \bm{\tilde{\gamma}_3} &\bm{\xi}\\
   \bm{0}  & \bm{0}  &\bm{\xi} & -i \bm{\tilde{\gamma}_4}\\
  \end{array}
 \right].
\label{28}   
\end{eqnarray}
The vectors $ \hat{\bm{a}} $ of field operators and $ \hat{\bm{L}}
$ of the Langevin operator forces in Eq.~(\ref{27}) are given as
$\hat{\bm{a}}=\left[\hat{\bm{a}}_{\bm{1}}, \hat{\bm{a}}_{\bm{2}},
\hat{\bm{a}}_{\bm{3}}, \hat{\bm{a}}_{\bm{4}}  \right]^T
\equiv\left[\hat{a}_{1},
\hat{a}_{1}^{\dagger},\hat{a}_{2},\hat{a}_{2}^{\dagger},\hat{a}_{3},\hat{a}_{3}^{\dagger},\hat{a}_{4},\hat{a}_{4}^{\dagger}
\right]^T$ and $\hat{\bm{L}}=\left[\hat{L}_{1},
\hat{L}_{1}^{\dagger},\hat{L}_{2},\hat{L}_{2}^{\dagger},\hat{L}_{3},\hat{L}_{3}^{\dagger},\hat{L}_{4},\hat{L}_{4}^{\dagger}
\right]^T$.

QEPs and QHPs have been found in the following two configurations:

\subsubsection{Linear configuration with equal damping and/or
amplification rates of neighbor modes}

In this configuration schematically shown in Fig.~\ref{fig1}(c) we
assume equal damping and/or amplification rates in modes 1 and 2,
and also modes 3 and 4:
\begin{equation}   
 \gamma_1 =\gamma_2 \equiv \gamma_{12}, \hspace{5mm}
 \gamma_3 =\gamma_4 \equiv \gamma_{34}.
\label{29}
\end{equation}
In this case, diagonalization of the dynamical  {$ 4\times 4 $}
matrix $\bm{M^{(4)}_{\rm l1} }$ in Eq.~(\ref{28}) provides the
eigenvalues
\begin{eqnarray}  
 \lambda_{1,2}^{M^{(4)}_{\rm l1}}  &=& -i \gamma_{+} \pm {\alpha}_{-}, \nonumber \\
 \lambda_{3,4}^{M^{(4)}_{\rm l1}}  &=& -i \gamma_{+} \pm
 {\alpha}_{+},
\label{30}   
\end{eqnarray}
and the corresponding eigenvectors
\begin{eqnarray}  
 \bm{y_{1,3}^{M^{(4)}_{\rm l1}} }& =& \left[-\frac{ \delta^*_{\pm} }{{2\xi^3}},
  \frac{\chi_{\mp}^{*2}}{{\xi^2}} -1,\frac{\chi_{\mp}^{*}}{{\xi}} , 1 \right]^T, \nonumber \\
 \bm{y_{2,4}^{M^{(4)}_{\rm l1}} }& =& \left[\frac{ \delta_{\pm} }{{2\xi^3}},
   \frac{\chi_{\mp}^{2}}{{\xi^2}} -1,-\frac{\chi_{\mp}}{{\xi}} , 1 \right]^T.\nonumber \\
\label{31}  
\end{eqnarray}
where $ \delta_{\pm} = \xi^{2} (-2i \gamma_{-} + \chi_{\mp} ) \pm
2 \mu \chi_{\mp} $,  $ \chi_{\pm}=-i \gamma_{+} + \alpha_{\pm} $,
$ \alpha_{\pm}^2= \beta^2 \pm \mu $, $ \beta^2= 3\xi^2/2 -
\gamma_{-}^2$, and $ \mu^2=4\xi^2 (5\xi^2/16 - \gamma_{-}^2) $
using $ 4\gamma_{\pm} = \gamma_{12} \pm \gamma_{34}$.

If $\mu=0$ then $ {\alpha}_{+}={\alpha}_{-}$ and
$\chi_{+}=\chi_{-}$. The eigenvalues in Eq.~(\ref{30}) are doubly
degenerated: $ \lambda_{1}^{M^{(4)}_{\rm l1}} =
\lambda_{3}^{M^{(4)}_{\rm l1}} $ and $ \lambda_{2}^{M^{(4)}_{\rm
l1}} = \lambda_{4}^{M^{(4)}_{\rm l1}} $. The same holds also for
their eigenvectors given in Eq.~(\ref{31}) and we have two QEPs
with second-order ED. Similarly as in the case of three-mode
bosonic system, the inclusion of the structure of submatrices in
the   {$ 8\times 8 $} matrix $ \bm{ M^{(4)}_{\rm l1} } $ gives
the second-order DD to these QEPs and we, thus, have two QHPs.
They are localized at the ellipse
\begin{equation}   
 \frac{\kappa^2}{\epsilon^2}+ \frac{16\gamma_{-}^2}{5\epsilon^2}=1
\label{32}
\end{equation}
defined in the parameter space $
(\kappa/\epsilon,\gamma_-/\epsilon) $. Real parts of the
eigenvalues $ \lambda_j^{M^{(4)}_{\rm l1} } $ for $ j=1,\ldots, 4
$ are drawn in this space in Fig.~\ref{fig3}(c);  {their
coinciding values for the condition (\ref{32}) are plotted by the
red dashed curves.}

The eigenvalues and eigenvectors of the $ 8\times 8 $ matrix $
\bm{ M^{(4)}_{\rm l1}} $ obtained by merging those written in
Eqs.~(\ref{30}) and (\ref{31}) and Eqs.~(\ref{7}) and (\ref{8})
for the matrix $ \bm{\xi} $ (for details, see Appendix~\ref{AppA})
 {using the scheme in Fig.~\ref{fig2}} define the
transformation into the new field operators
$\hat{\bm{b}}=\left[\hat{b}_{1},\hat{b}_{2},\hat{b}_{1}^{\dagger},
\hat{b}_{2}^{\dagger},\hat{b}_{3},\hat{b}_{4},\hat{b}_{3}^{\dagger},\hat{b}_{4}^{\dagger}
\right]^T$ in which the Heisenberg-Langevin equations have the
diagonal form. This reveals the spectral degeneracies in
higher-order FOMs that are summarized in Tab.~\ref{tab3} in
Appendix~\ref{AppB} considering the first- and second-order FOMs.

\subsubsection{Linear configuration with different damping and/or amplification rates for neighbor modes}

We assume the equal damping and/or amplification rates in modes 1
and 3 and also in modes 2 and 4:
\begin{equation}   
 \gamma_1=\gamma_3\equiv\gamma_{13}, \hspace{5mm}
 \gamma_2=\gamma_4\equiv\gamma_{24}.
\label{33}
\end{equation}
In this configuration, which is schematically depicted in
Fig.~\ref{fig1}(e), the eigenvalues of the $ 4\times 4 $ dynamical
matrix $\bm{M^{(4)}_{\rm l2}}$ are obtained as follows:
\begin{eqnarray} 
 \lambda_{1,2}^{M^{(4)}_{\rm l2}}  &=& -i \gamma_{+}\pm \alpha_{+}, \nonumber \\
 \lambda_{3,4}^{M^{(4)}_{\rm l2}}  &=& -i \gamma_{+} \pm
 \alpha_{-}.
\label{34}  
\end{eqnarray}
The corresponding eigenvectors are reached in the form:
\begin{eqnarray}  
 \bm{y_{1,3}^{M^{(4)}_{\rm l2}} } &=& \left[\frac{ -\kappa_{\mp} }{{\xi}},
  \kappa_{\pm},\frac{\chi_{\pm}^{*}}{{\xi}} , 1 \right]^T, \nonumber \\
 \bm{y_{2,4}^{M^{(4)}_{\rm l2}} } &=& \left[\frac{ \kappa_{\mp} \chi_{\pm}}{{\xi}}, \kappa_{\pm} ,-\frac{\chi_{\pm}}{{\xi}} , 1 \right]^T,\nonumber \\
\label{35}   
\end{eqnarray}
where $ \kappa_{\pm} = (\pm\sqrt{5}+1)/2 $, ${\chi}_{\pm}=-i
\gamma_{+} + \alpha_{\pm} $, $\alpha_{\pm}^2= \tilde{\beta} \pm
\mu$, $ \tilde{\beta}= 3\xi^2/2 - \gamma_{-}^2$, and $ \mu=
\sqrt{5}\xi^2/2 $ using $ 4\gamma_{\pm} = \gamma_{13} \pm
\gamma_{24}$.

Provided that $\alpha_{+}=0$ [$\alpha_{-}=0$] the eigenvalues $
\lambda_{1}^{M^{(4)}_{\rm l2}} $ and $ \lambda_{2}^{M^{(4)}_{\rm
l2}} $ [$ \lambda_{3}^{M^{(4)}_{\rm l2}} $ and $
\lambda_{4}^{M^{(4)}_{\rm l2}} $] in Eq.~(\ref{34}) and the
eigenvectors $ \bm{y_{1}^{M^{(4)}_{\rm l2}}} $ and $
\bm{y_{2}^{M^{(4)}_{\rm l2}}} $ [$ \bm{y_{3}^{M^{(4)}_{\rm l2}}} $
and $ \bm{y_{4}^{M^{(4)}_{\rm l2}}} $] in Eq.~(\ref{35}) are equal
to each other and we have a QEP with second-order ED. This means
that, for the $ 8\times 8 $ dynamical matrix $\bm{M^{(4)}_{\rm
l2}}$, we predict a QHP with second-order ED and DD observed
either for $\alpha_{+}=0$ or $\alpha_{-}=0$. These conditions
define two ellipses in the parameter space $
(\kappa/\epsilon,\gamma_-/\epsilon) $,
\begin{equation}   
 \frac{\kappa^2}{\epsilon^2}+ \frac{2\gamma_{-}^2}{ (3\pm \sqrt{5})\epsilon^2 }
  =1.
\label{36}
\end{equation}
They are shown in Fig.~\ref{fig3}(d) where the real parts of the
eigenvalues $ \lambda_j^{M^{(4)}_{\rm l2} } $ for $ j=1,\ldots, 4
$ are plotted;  {they are drawn by the red dashed curves for
the condition in Eq.~(\ref{36}).}

The eigenvalues and eigenvectors of the $ 8\times 8 $ matrix $
\bm{ M^{(4)}_{\rm l2}} $, given in Appendix~\ref{AppA}, determine
the transformation that leaves the Heisenberg-Langevin equations
in their diagonal form. Thus, we define the new field operators
$\hat{\bm{b}}=\left[\hat{b}_{1},\hat{b}_{2},\hat{b}_{1}^{\dagger},
\hat{b}_{2}^{\dagger},\hat{b}_{3},\hat{b}_{4},\hat{b}_{3}^{\dagger},\hat{b}_{4}^{\dagger}
\right]^T$, whose evolution allows to identify the QEPs and QHPs
in the first- and second-order FOM dynamics. They are given in
Tab.~\ref{tab4} of Appendix~\ref{AppB}.

\subsection{Circular configuration}

In the circular configuration, the Hamiltonian $ \hat{H}_{4,{\rm
c}} $ of four-mode system is expressed as
\begin{eqnarray}  
 \hat{H}_{4,{\rm c}} &=&  \left[ \hbar \epsilon \hat{a}_{1}^{\dagger}\hat{a}_{2}
   + \hbar \epsilon \hat{a}_{2}^{\dagger}\hat{a}_{3}
   + \hbar \epsilon \hat{a}_{3}^{\dagger}\hat{a}_{4}
   + \hbar \epsilon \hat{a}_{4}^{\dagger}\hat{a}_{1}  + \hbar \kappa \hat{a}_{1}\hat{a}_{2} \right.\nonumber \\
  & & \left. + \hbar \kappa \hat{a}_{2}\hat{a}_{3} + \hbar \kappa
   \hat{a}_{3}\hat{a}_{4}  +\hbar \kappa \hat{a}_{4}\hat{a}_{1} \right]  +
   \textrm{H.c.}
\label{37}   
\end{eqnarray}
Whereas the Heisenberg-Langevin equations keep the form of
Eq.~(\ref{27}), the original dynamical  {$ 4\times 4 $} matrix
$ \bm{M^{(4)}_{\rm l}} $ of the linear configuration in
Eq.~(\ref{28}) is modified into the form
\begin{eqnarray}  
 \bm{M^{(4)}_{\rm c}}  & = & \left[
 \begin{array}{cccc}
  -i \bm{\tilde{\gamma}_1} & \bm{\xi} & \bm{0} & \bm{\xi}\\
  \bm{\xi} & -i \bm{\tilde{\gamma}_2} & \bm{\xi}& \bm{0}\\
 \bm{0}  & \bm{\xi} & -i \bm{\tilde{\gamma}_3} &\bm{\xi}\\
 \bm{\xi}  & \bm{0}  &\bm{\xi} & -i \bm{\tilde{\gamma}_4}\\
 \end{array} \right].
\label{38}
\end{eqnarray}

We derive the eigenvalues of the  {$ 4\times 4 $} matrix $
\bm{M^{(4)}_{\rm c}} $ under the condition of the equal damping
and/or amplification rates of the neighbor modes valid for the
configuration plotted in Fig.~\ref{fig1}(d); i.e.,
\begin{equation}   
 \gamma_1 =\gamma_2 \equiv \gamma_{12}, \hspace{5mm}
 \gamma_3 =\gamma_4 \equiv \gamma_{34}.
\label{39}
\end{equation}
We obtain the following eigenvalues
\begin{eqnarray}  
 \lambda_{1,2}^{M^{(4)}_{\rm c1}} &=& -i
   \gamma_{+} - \alpha_{\pm}, \nonumber \\
 \lambda_{3,4}^{M^{(4)}_{\rm c1}} &=& -i \gamma_{+}
    +\alpha_{\mp}
\label{40}   
\end{eqnarray}
that are accompanied by the following eigenvectors
\begin{eqnarray}   
 \bm{y_{1,2}^{M^{(4)}_{\rm c1}} } &=& \left[-\frac{\chi}{\xi}, \pm \frac{\chi}{\xi} ,\mp 1 , 1 \right]^T, \nonumber \\
 \bm{y_{3,4}^{M^{(4)}_{\rm c1}} } &=& \left[\frac{\chi^{*}}{\xi}, \mp\frac{\chi^{*}}{\xi} ,\mp 1, 1
 \right]^T.
\label{41}   
\end{eqnarray}
The symbols introduced in Eqs.~(\ref{40}) and (\ref{41}) are
defined as $ \chi=i \gamma_{+} + \beta $, $ \alpha_{\pm}= \beta
\pm \xi $, $ \beta^2= \xi^2- \gamma_{-}^2 $, and $ 4\gamma_{\pm} =
\gamma_{12} \pm \gamma_{34} $.

For $ \beta=0 $, we have $ \alpha_{+}=-\alpha_{-}$ and $
\chi^*=-\chi $. This leads to the relations $
\lambda_{1}^{M^{(4)}_{\rm c1}} = \lambda_{3}^{M^{(4)}_{\rm c1}} $
and  $ \lambda_{2}^{M^{(4)}_{\rm c1}} = \lambda_{4}^{M^{(4)}_{\rm
c1}} $. Also the corresponding eigenvectors coincide: $
\bm{y_{1}^{M^{(4)}_{\rm c1}} } = \bm{y_{3}^{M^{(4)}_{\rm c1}} } $
and $ \bm{y_{2}^{M^{(4)}_{\rm c1}} } = \bm{y_{4}^{M^{(4)}_{\rm
c1}} } $. As a consequence, the two QEPs with second-order EDs
occur. This means that the two QHPs with second-order EDs and DDs
for the $ 8\times 8 $ matrix $ \bm{M^{(4)}_{\rm c1}} $ are formed.
These QHPs occur under the condition specified in Eq.~(\ref{13})
that identifies the QHPs in the analyzed two-mode bosonic system.
The real values of the eigenvalues $ \lambda_{j}^{M^{(4)}_{\rm
c1}} $ for $ j=1,\ldots, 4 $ are plotted in Fig.~\ref{fig3}(e)
together with the condition for the QHPs;  {the red dashed
curves give the coinciding values under the condition in
Eq.~(\ref{13}).}

Using the eigenvalues and eigenvectors of the $ 4\times 4 $ matrix
$ \bm{M^{(4)}_{\rm c1}} $ and the $ 2\times 2 $ matrix $ \bm{\xi}
$ given in Eqs.~(\ref{40}), (\ref{41}), (\ref{7}), and (\ref{8}),
we arrive at the system dynamical matrix in its diagonal form (for
details, see Appendix~\ref{AppA}). This brings us the new field
operators $
\hat{\bm{b}}=\left[\hat{b}_{1},\hat{b}_{2},\hat{b}_{3}^{\dagger},
\hat{b}_{4}^{\dagger},\hat{b}_{3},\hat{b}_{4},\hat{b}_{1}^{\dagger},
\hat{b}_{2}^{\dagger} \right]^T $ suitable for revealing QEPs and
QHPs found in the dynamics of FOMs. As the structure of
eigenvalues and eigenvectors is the same as that characterizing
the four-mode linear system with equal damping and/or
amplification rates of neighbor modes, the appropriate QEPs and
QHPs are given in Tab.~\ref{tab3} in Appendix~\ref{AppB}.

\section{Five-mode bosonic systems}

The largest bosonic systems in our investigations consist of five
bosonic modes. Among them, QEPs and QHPs were identified in two
configurations: linear and pyramid.

\subsection{Linear configuration}

A five-mode bosonic system in its linear configuration is depicted
in Fig.~\ref{fig1}(f). Its Hamiltonian $ \hat{H}_{5,{\rm l}} $ is
given as follows:
\begin{eqnarray}   
 \hat{H}_{5,{\rm l}} &=& \Bigl[ \hbar \epsilon \hat{a}_{1}^{\dagger}\hat{a}_{2}
  + \hbar \epsilon \hat{a}_{2}^{\dagger}\hat{a}_{3}
  + \hbar \epsilon \hat{a}_{3}^{\dagger}\hat{a}_{4}
  + \hbar \epsilon \hat{a}_{4}^{\dagger}\hat{a}_{5} + \hbar \kappa \hat{a}_{1}\hat{a}_{2}\nonumber \\
 & & + \hbar \kappa \hat{a}_{2}\hat{a}_{3} + \hbar \kappa \hat{a}_{3}\hat{a}_{4}
 + \hbar \kappa \hat{a}_{4}\hat{a}_{5} \Bigr] +\textrm{H.c.}
\label{42}    
\end{eqnarray}
Defining the vectors $\hat{\bm{a}}=\left[\hat{\bm{a}}_{\bm{1}},
 \hat{\bm{a}}_{\bm{2}},  \hat{\bm{a}}_{\bm{3}},
\hat{\bm{a}}_{\bm{4}},  \hat{\bm{a}}_{\bm{5}}  \right]^T
\equiv\left[\hat{a}_{1},
\hat{a}_{1}^{\dagger},\hat{a}_{2},\hat{a}_{2}^{\dagger},\hat{a}_{3},\hat{a}_{3}^{\dagger},\hat{a}_{4},\hat{a}_{4}^{\dagger},\hat{a}_{5},\hat{a}_{5}^{\dagger}
\right]^T$ of field operators and $\hat{\bm{L}}=\left[\hat{L}_{1},
\hat{L}_{1}^{\dagger},\hat{L}_{2},\hat{L}_{2}^{\dagger},\hat{L}_{3},\hat{L}_{3}^{\dagger},\hat{L}_{4},\hat{L}_{4}^{\dagger}
,\hat{L}_{5},\hat{L}_{5}^{\dagger} \right]^T$ of the Langevin
operator forces, we obtain the following Heisenberg-Langevin
equations:
\begin{eqnarray}    
 \frac{d\hat{\bm{a}}}{dt} & = & -i \bm{M^{(5)}_{\rm l}} \hat{\bm{a}}
 +\hat{\bm{L}}.
\label{43}   
\end{eqnarray}
The dynamical  {$ 10\times 10 $} matrix $ \bm{M^{(5)}_{\rm l}}
$ introduced in Eq.~(\ref{43}) is written with the help of $
2\times 2 $ submatrices introduced in Eq.~(\ref{4}) as
\begin{eqnarray}   
 \bm{M^{(5)}_{\rm l}}  & = & \left[
 \begin{array}{ccccc}
  -i \bm{\tilde{\gamma}_1} & \bm{\xi} & \bm{0} & \bm{0}& \bm{0}\\
  \bm{\xi} & -i \bm{\tilde{\gamma}_2} & \bm{\xi}& \bm{0}& \bm{0}\\
 \bm{0}  & \bm{\xi} & -i \bm{\tilde{\gamma}_3} &\bm{\xi}& \bm{0}\\
 \bm{0}  & \bm{0}  &\bm{\xi} & -i \bm{\tilde{\gamma}_4}& \bm{\xi}\\
 \bm{0}  & \bm{0}  &\bm{0} & \bm{\xi} & -i \bm{\tilde{\gamma}_5}\\
 \end{array} \right].
\label{44}    
\end{eqnarray}

Motivated by $ \mathcal{PT} $ symmetry we assume:
\begin{equation}   
 \gamma_1 =\gamma_2\equiv \gamma_{12}, \hspace{5mm}
 \gamma_4 =\gamma_5\equiv \gamma_{45}.
\label{45}
\end{equation}
Moreover, inspired by the condition in Eq.~(\ref{22}) derived for
the three-mode linear system, we additionally assume:
\begin{equation}   
2 \gamma_3 =\gamma_{12}+\gamma_{45}.
 \label{46}
\end{equation}
Then, diagonalization of the  {$ 5\times 5 $} matrix $
\bm{M^{(5)}_{\rm l}} $ results in the following eigenvalues $
{\lambda_{j}^{M^{(5)}_{\rm l}} } $:
\begin{eqnarray}  
 \lambda_1^{M^{(5)}_{\rm l}}  &=& -i \gamma_{+}, \nonumber \\
 \lambda_{2,3}^{M^{(5)}_{\rm l}}  &=& -i \gamma_{+} \pm \alpha_{-},\nonumber \\
 \lambda_{4,5}^{M^{(5)}_{\rm l}}  &=& -i \gamma_{+} \pm \alpha_{+}.
\label{47}   
\end{eqnarray}
The corresponding eigenvectors are derived as follows:
\begin{eqnarray}  
 \bm{y_{1}^{M^{(5)}_{\rm l}} } &=& \left[1, \frac{i \gamma_{-}}{\xi}, -\frac{\gamma_{-}^2}{\xi^2} -1,-\frac{i \gamma_{-}}{\xi} , 1 \right]^T, \nonumber \\
 \bm{y_{2,4}^{M^{(5)}_{\rm l}} } &=& \left[-1\mp\frac{n_{\mp}}{4\xi^3}, -\frac{m_{\mp} }{\xi^3} \mp \frac{2 \beta \chi_{\mp} }{\xi^2},
   \frac{\chi_{\mp}^{2}}{{\xi^2}} -1,\frac{\chi_{\mp}}{\xi} , 1 \right]^T, \nonumber \\
 \bm{y_{3,5}^{M^{(5)}_{\rm l}} }& =&\left[-1\mp\frac{n_{\mp}^{*}}{4\xi^3}, \frac{m_{\mp}^{*} }{\xi^3} \pm \frac{2 \beta \chi_{\mp}^{*} }{\xi^2},
   \frac{\chi_{\mp}^{*2}}{{\xi^2}} -1,-\frac{\chi_{\mp}^{*}}{{\xi}} , 1 \right]^T,\nonumber \\
\label{48}   
\end{eqnarray}
where $ n_{\mp}= (2\beta\mp\xi) [ (2\beta\mp\xi)^2-4i \gamma_{-}
\alpha_{\mp} ]$, $ m_{\mp}= i\gamma_{-} (\chi_{\mp}^{2} +\xi^2 )$,
${\chi}_{\pm}=-i \gamma_{+} + \alpha_{\pm} $, $ \alpha_{\pm}^2=
\beta^2+ 7\xi^2/4 \pm 2\beta\xi $, $ \beta^2= \xi^2/4
-\gamma_{-}^2 $, and $ 4 \gamma_{\pm} = \gamma_{12} \pm
\gamma_{45} $.

If $\beta=0$ then $ \alpha_{+}=\alpha_{-}$ and
$\chi_{+}=\chi_{-}$. Under these conditions, we have $
\lambda_{2}^{M^{(5)}_{\rm l}} = \lambda_{4}^{M^{(5)}_{\rm l}} $
and $ \bm{y_{2}^{M^{(5)}_{\rm l}} } = \bm{y_{4}^{M^{(5)}_{\rm l}}
} $. We also have $ \lambda_{3}^{M^{(5)}_{\rm l}} =
\lambda_{5}^{M^{(5)}_{\rm l}} $ and $ \bm{y_{3}^{M^{(5)}_{\rm l}}
} = \bm{y_{5}^{M^{(5)}_{\rm l}} } $. Thus, we observe two QEPs
with second-order EDs that give rise to two QHPs with second-order
EDs and DDs when the $ 10\times 10 $ matrix $ \bm{M^{(5)}_{\rm l}}
$ is analyzed. In the parameter space $
(\kappa/\epsilon,\gamma_-/\epsilon) $, these QHPs are localized at
the positions obeying the condition
\begin{equation}   
 \frac{\kappa^2}{\epsilon^2}+ \frac{4\gamma_{-}^2}{\epsilon^2}=1.
\label{49}
\end{equation}
The real parts of the eigenvalues $ \lambda_j^{M^{(5)}_{\rm l}} $
forming QHPs ($ j=2,\ldots, 5 $) are plotted in Fig.~\ref{fig3}(f)
 {where their coinciding values for the condition in
Eq.~(\ref{49}) are indicated by the red dashed curves.}

The eigenvalues and eigenvectors of the $ 10\times 10 $ matrix $
\bm{ M^{(5)}_{\rm l} } $ constructed from the formulas given in
Eqs.~(\ref{47}) and (\ref{48}) together with those in
Eqs.~(\ref{7}) and (\ref{8}) (for details, see
Appendix~\ref{AppA}) allow us to describe the system dynamics via
a diagonal dynamical matrix. The appropriate operators
$\hat{\bm{b}}=\left[\hat{b}_{1},\hat{b}_{1}^{\dagger},\hat{b}_{2},\hat{b}_{3},
\hat{b}_{2}^{\dagger},\hat{b}_{3}^{\dagger},\hat{b}_{4},\hat{b}_{5},\hat{b}_{4}^{\dagger},\hat{b}_{5}^{\dagger}
\right]^T$ then reveal QEPs and QHPs found in the dynamics of FOMs
of different orders. For the first- and second-order FOMs they are
written in Tab.~\ref{tab5} of Appendix~\ref{AppB}.

\subsection{Pyramid configuration}

Second-order QEPs and QHPs can also be identified in the pyramid
configuration [the only considered non-planar configuration, see
Fig.~\ref{fig1}(g)] described by the following Hamiltonian $
\hat{H}_{5,{\rm p}} $:
\begin{eqnarray}   
 \hat{H}_{5,{\rm p}} &=& \left[ \hbar \epsilon \hat{a}_{1}^{\dagger}\hat{a}_{2} + \hbar \epsilon
 \hat{a}_{1}^{\dagger}\hat{a}_{4} + \hbar \epsilon \hat{a}_{1}^{\dagger}\hat{a}_{5}
  + \hbar \epsilon \hat{a}_{2}^{\dagger}\hat{a}_{3} + \hbar \epsilon \hat{a}_{2}^{\dagger}\hat{a}_{5} \right.   + \hbar \epsilon \hat{a}_{3}^{\dagger}\hat{a}_{4}
  + \hbar \epsilon \hat{a}_{3}^{\dagger}\hat{a}_{5}
    \nonumber \\
 & & + \hbar \epsilon \hat{a}_{4}^{\dagger}\hat{a}_{5} + \hbar \kappa \hat{a}_{1}\hat{a}_{2} + \hbar \kappa \hat{a}_{1}\hat{a}_{4} + \hbar \kappa \hat{a}_{1}\hat{a}_{5}  + \hbar \kappa \hat{a}_{2}\hat{a}_{3} + \hbar \kappa \hat{a}_{2}\hat{a}_{5} \nonumber \\
 & & \left.  + \hbar \kappa \hat{a}_{3}\hat{a}_{4}
  + \hbar \kappa \hat{a}_{3}\hat{a}_{5} + \hbar \kappa \hat{a}_{4}\hat{a}_{5}
  \right] +\textrm{H.c.}
\label{50}    
\end{eqnarray}

The Hamiltonian $ \hat{H}_{5,{\rm p}} $ given in Eq.~(\ref{50})
leads to the Heisenberg-Langevin equations (\ref{43}) in which the
dynamical  {$ 10\times 10 $} matrix $ \bm{M^{(5)}_{\rm p}} $
attains the form using the $ 2\times 2 $ submatrices introduced in
Eq.~(\ref{4}):
\begin{eqnarray}   
 \bm{M^{(5)}_{\rm p}}  & = & \left[
 \begin{array}{ccccc}
  -i \bm{\tilde{\gamma}_1} & \bm{\xi} & \bm{0} & \bm{\xi}& \bm{\xi}\\
  \bm{\xi} & -i \bm{\tilde{\gamma}_2} & \bm{\xi}& \bm{0}& \bm{\xi}\\
 \bm{0}  & \bm{\xi} & -i \bm{\tilde{\gamma}_3} &\bm{\xi}& \bm{\xi}\\
 \bm{\xi}  & \bm{0}  &\bm{\xi} & -i \bm{\tilde{\gamma}_4}& \bm{\xi}\\
 \bm{\xi}  & \bm{\xi}  &\bm{\xi} & \bm{\xi} & -i \bm{\tilde{\gamma}_5}\\
 \end{array} \right].
\label{51}    
\end{eqnarray}

Motivated by the $ \mathcal{PT} $ symmetry we assume:
\begin{equation}   
 \gamma_1 =\gamma_2\equiv \gamma_{12}, \hspace{5mm}
 \gamma_3 =\gamma_4\equiv \gamma_{34}.
\label{52}
\end{equation}
Moreover, inspired by the condition (\ref{22}) derived for the
three-mode linear bosonic system, we additionally assume:
\begin{equation}   
2 \gamma_5 =\gamma_{12}+\gamma_{34}.
 \label{53}
\end{equation}

Under these conditions, diagonalization of the   {$ 5\times 5$}
matrix $ \bm{M^{(5)}_{\rm p}} $ leaves us the following five
eigenvalues $ {\lambda_{j}^{M^{(5)}_{\rm p}} } $:
\begin{eqnarray}  
 \lambda_1^{M^{(5)_{\rm p}}}  &=& -i \gamma_{+}, \nonumber \\
 \lambda_{2,3}^{M^{(5)}_{\rm p}}  &=& -i \gamma_{+} -\xi \pm \beta_1,\nonumber \\
 \lambda_{4,5}^{M^{(5)}_{\rm p}}  &=& -i \gamma_{+} +\xi \pm
 \beta_2.
\label{54}   
\end{eqnarray}
The corresponding eigenvectors are obtained as follows:
\begin{eqnarray}  
 \bm{y_{1}^{M^{(5)}_{\rm p}} } &=& \left[\frac{-i \xi}{\gamma_{-}}, \frac{-i \xi}{\gamma_{-}}, \frac{i \xi}{\gamma_{-}},\frac{i \xi}{\gamma_{-}} , 1 \right]^T, \nonumber \\
 \bm{y_{2}^{M^{(5)}_{\rm p}} } &=& \left[\frac{\chi_{1}}{\xi}, -\frac{\chi_{1}}{\xi}, -1, 1, 0 \right]^T, \nonumber \\
 \bm{y_{3}^{M^{(5)}_{\rm p}} } &=& \left[-\frac{\chi_{1}^{*}}{\xi}, \frac{\chi_{1}^{*}}{\xi}, -1, 1, 0 \right]^T, \nonumber \\
 \bm{y_{4}^{M^{(5)}_{\rm p}} }& =& \left[ \sigma_+, \sigma_+,
 \sigma_+^*, \sigma_+^* , 1 \right]^T,\nonumber \\
 \bm{y_{5}^{M^{(5)}_{\rm p}} }& =& \left[ \sigma_-^*, \sigma_-^*,
 \sigma_-, \sigma_- , 1 \right]^T,\nonumber \\
\label{55}   
\end{eqnarray}
where $ \sigma_{\pm} = (\xi \pm \chi_{2}) /(4\xi) $, $ \beta_1^2=
\xi^2 -\gamma_{-}^2 $, $ \beta_2^2= 5\xi^2 -\gamma_{-}^2 $,
$\chi_{1,2}=\beta_{1,2}-i \gamma_{-}$, and $ 4 \gamma_{\pm} =
\gamma_{12} \pm \gamma_{34} $.

Contrary to the eigenvalues of the models discussed above, the
eigenvalues $ \lambda_j^{M^{(5)}_{\rm p}} $ for $ j=2,\ldots, 5 $
in Eq.~(\ref{55}) exhibit the linear dependence on $ \xi $. This
means that the DD inherited to all the above-discussed models is
removed in this model. It remains only for the eigenvalue $
\lambda_1^{M^{(5)}_{\rm p}} $ that, however, has no ability to
form EDs.

Provided that $\beta_1=0$, we have $ \lambda_{2}^{M^{(5)}_{\rm p}}
= \lambda_{3}^{M^{(5)}_{\rm p}} $ and $ \bm{y_{2}^{M^{(5)}_{\rm
p}} } = \bm{y_{3}^{M^{(5)}_{\rm p}} } $. Thus, we have the QEP
with second-order ED. When the $ 10\times 10 $ matrix $
\bm{M^{(5)}_{\rm p}} $ is analyzed, there occur one second-order
QEP for $ \xi = \zeta $ and one second-order QEP for $ \xi = -
\zeta $. These QEPs occur in the parameter space $
(\kappa/\epsilon,\gamma_-/\epsilon) $ under the condition written
in Eq.~(\ref{13}).

Moreover, in parallel, if $\beta_2=0$, it holds that $
\lambda_{4}^{M^{(5)}_{\rm p}} = \lambda_{5}^{M^{(5)}_{\rm p}} $
and $ \bm{y_{4}^{M^{(5)}_{\rm p}} } = \bm{y_{5}^{M^{(5)}_{\rm p}}
} $. Thus, similarly as above, we observe the QEP with
second-order ED for the $ 5\times 5 $ matrix $ \bm{M^{(5)}_{\rm
p}} $. There occur one second-order QEP for $ \xi = \zeta $ and
one second-order QEP for $ \xi = -\zeta $ for the $ 10\times 10 $
matrix $ \bm{M^{(5)}_{\rm p}} $. These QEPs are localized in the
parameter space $ (\kappa/\epsilon,\gamma_-/\epsilon) $ at the
points fulfilling the condition:
\begin{equation}   
 \frac{\kappa^2}{\epsilon^2}+ \frac{\gamma_{-}^2}{5\epsilon^2}=1.
\label{56}
\end{equation}
The real parts of the eigenvalues $ \lambda_j^{M^{(5)}_{\rm p}} $
forming QEPs ($ j=2,\ldots, 5 $) are drawn in
Fig.~\ref{fig3}(g,h);  { the coinciding values occurring for
the condition in Eq.~(\ref{13}) [(\ref{56})] are plotted in
Fig.~\ref{fig3}(g) [(h)] as red dashed curves.}

We note that for $ \xi = 0 $ (i.e., $ \gamma_- = 0 $, $ \kappa =
\epsilon $) we have a specific QHP with ten-fold frequency
degeneracy that belongs to four doubly-degenerate eigenvectors and
another two eigenvectors.

The eigenvalues and eigenvectors of the $ 10\times 10 $ matrix $
\bm{ M^{(5)}_{\rm p} } $ formed from the expressions given in
Eqs.~(\ref{54}) and (\ref{55}) and also in Eqs.~(\ref{7}) and
(\ref{8}) (for details, see Appendix~\ref{AppA}) using the scheme
in Fig.~\ref{fig2} allow us to analyze the system dynamics via a
diagonal dynamical matrix. The appropriate operators
$\hat{\bm{b}}=\left[\hat{b}_{1},\hat{b}_{1}^{\dagger},\hat{b}_{2},\hat{b}_{3},
\hat{b}_{3}^{\dagger},\hat{b}_{2}^{\dagger},\hat{b}_{4},\hat{b}_{5},\hat{b}_{5}^{\dagger},\hat{b}_{4}^{\dagger}
\right]^T$ then allow to identify the QEPs and QHPs observed in
the dynamics of FOMs of different orders. For the first- and
second-order FOMs, they are explicitly given in Tab.~\ref{tab6} in
Appendix~\ref{AppB}.

\section{Conclusions}
\label{sec7}

We have analyzed the dynamics of simple bosonic systems described
by quadratic non-Hermitian Hamiltonians from the point of view of
the occurrence of quantum exceptional, diabolical, and hybrid
points. Non-Hermiticity of the considered systems, composed of
from two to five coupled modes, originated in their damping and/or
amplification, that are accompanied by the corresponding Langevin
fluctuating forces to assure the physically consistent behavior.
We have identified specific configurations defined by two-mode
couplings and conditions for the damping and amplification rates
of the modes at which the inherited quantum exceptional and
diabolical points occur. Surprisingly, in these physically
consistent models, we have found only second- and third-order
inherited quantum exceptional points including their doubling due
to second-order diabolical degeneracies.  {On the other hand,
doubled second-order inherited quantum exceptional points have
been observed.} We have shown that the analyzed bosonic systems
naturally exhibit the second-order diabolical degeneracies.
 {Nevertheless, we have found an exception from this behavior
in which no diabolical degeneracy occurs.}

The exceptional and diabolical degeneracies of inherited quantum
hybrid points have then been used to construct higher-order
degeneracies observed in the dynamics of second-order
field-operator moments. The corresponding quantum exceptional and
hybrid points are summarized in tables that demonstrate richness
of the evolution of the general-order field-operator moments and
 {serve for future studies of the related physical effects.}

The investigations have revealed the need for further looking for
the bosonic systems exhibiting higher-order inherited quantum
exceptional and hybrid points by considering more general bosonic
systems. In Ref.~\cite{PerinaJr2025} we extend our investigations
to the systems with partial $ \mathcal{PT} $-symmetry like
dynamics (nonconventional $ \mathcal{PT} $-symmetry) as well as
non-Hermitian bosonic systems with unidirectional coupling.

We may conclude in general that the performed analysis opens the
door for further detailed investigations of the role of
exceptional and diabolical degeneracies responsible for inducing
unusual physical effects observed in physically well-behaved
systems at exceptional, diabolical, and hybrid points.

\section{Acknowledgements}

The authors thank Ievgen I. Arkhipov for useful discussions. J.P.
and K.T. acknowledge support by the project OP JAC
CZ.02.01.01/00/22\_008/0004596 of the Ministry of Education,
Youth, and Sports of the Czech Republic. J.P. acknowledges support
by the project No. 25-15775S of the Czech Science Foundation.
A.K.-K., G.Ch., and A.M. were supported by the Polish National
Science Centre (NCN) under the Maestro Grant No.
DEC-2019/34/A/ST2/00081.

\appendix
\section{Eigenvalues and eigenvectors of three-, four- and five-mode bosonic
systems}  
\label{AppA}

In this Appendix, we present the eigenvalues and eigenvectors of
the dynamical matrices of the Heisenberg-Langevin equations for
the three-, four-, and five-mode bosonic systems in configurations
depicted in Fig.~\ref{fig1}.

\subsection{General $ n $-mode bosonic systems}

The eigenvalues of the $ 2n \times 2n $ dynamical matrix
$\bm{M^{(n)}}$ belonging to a system with $ n $ modes ($
n=2,3,\ldots $) are constructed from the eigenvalues $
\lambda_j^{M^{(n)}} $, $ j=1,\ldots,n $, derived for the form of
the dynamical matrix containing $ 2\times 2 $ submatrices and the
eigenvalues $ \lambda^\xi_j $, $ j=1,2 $, of the matrix $ \bm{\xi}
$ given in Eq.~(\ref{7}) as follows:
\begin{eqnarray}   
 \Lambda_{2j-1}^{M^{(n)}} &=&
   {\lambda_{j}^{M^{(n)}}}({\xi}=\lambda^{\xi}_{1}),
   \nonumber \\
 \Lambda_{2j}^{M^{(n)}} &=& {\lambda_{j}^{M^{(n)}} }
   ({\xi}=\lambda^{\xi}_{2}),
\hspace{5mm} j=1,2,\ldots,n. \label{A1}
\end{eqnarray}

Similarly, using the corresponding eigenvectors $
\bm{y_{j}^{M^{(n)}} } $ of the $ n\times n $ matrix $\bm{M^{(n)}}$
formed by submatrices and the eigenvectors $ \bm{y^{\xi}_{j}} $, $
j=1,2 $, of the matrix $ \bm{\xi} $ in Eq.~(\ref{8}), we arrive at
the following eigenvectors of the $ 2n\times 2n $ matrix
$\bm{M^{(n)}} $ associated with the eigenvalues given in
Eq.~(\ref{A1}):
\begin{eqnarray}   
 \bm{Y_{2j-1}^{M^{(n)}}} &=& \left[
  \begin{array}{c}
   {y_{j,1}^{M^{(n)}} }({\xi}=\lambda^{\xi}_{1}) \bm{y^{\xi}_{1}}  \\
   {y_{j,2}^{M^{(n)}} }({\xi}=\lambda^{\xi}_{1}) \bm{y^{\xi}_{1}}\\
   \ldots \\
   {y_{j,n}^{M^{(n)}} }({\xi}=\lambda^{\xi}_{1}) \bm{y^{\xi}_{1}} \end{array}\right],
   \nonumber \\
 \bm{Y_{2j}^{M^{(n)}}} &=& \left[
  \begin{array}{c}
   {y_{j,1}^{M^{(n)}} }({\xi}=\lambda^{\xi}_{2}) \bm{y^{\xi}_{2}}  \\
   {y_{j,2}^{M^{(n)}} }({\xi}=\lambda^{\xi}_{2}) \bm{y^{\xi}_{2}}\\
   \ldots \\
   {y_{j,n}^{M^{(n)}} }({\xi}=\lambda^{\xi}_{2}) \bm{y^{\xi}_{2}} \end{array}\right],
 \hspace{2mm} j=1,\ldots,n.
\label{A2}
\end{eqnarray}

\subsection{Three-mode bosonic system}

In the three-mode system ($ n=3 $) assuming $\beta=
\sqrt{2\zeta^2- \gamma_{-}^2} = 0$, we find a QHP with third-order
ED and second-order DD. All the eigenvalues in Eq.~(\ref{A1}) are
the same in this case and the eigenvectors in Eq.~(\ref{A2}) obey
the relations
$\bm{Y_{1}^{M^{(3)}}}=\bm{Y_{3}^{M^{(3)}}}=\bm{Y_{5}^{M^{(3)}}}$
and
$\bm{Y_{2}^{M^{(3)}}}=\bm{Y_{4}^{M^{(3)}}}=\bm{Y_{6}^{M^{(3)}}}$.

\subsection{Four-mode bosonic systems}

In the four-mode system ($ n=4 $) in the linear configuration with
equal damping and/or amplification rates of neighbor modes and
assuming $ \mu= 2\sqrt{ \zeta^2 (5\zeta^2/16  - \gamma_{-}^2) } =
0 $, we observe two QHPs with second-order ED and second-order DD.
The eigenvalues $ \Lambda_{1,2,5,6}^{M^{(4)}_{\rm l1}} $ and also
the eigenvalues $ \Lambda_{3,4,7,8}^{M^{(4)}_{\rm l1}} $ in
Eq.~(\ref{A1}) coincide. The eigenvectors in Eq.~(\ref{A2}) obey
the relations $\bm{ Y_{j}^{M^{(4)}_{\rm l1}} }= \bm{
Y_{j+4}^{M^{(4)}_{\rm l1}} } $ for $ j=1,\ldots,4 $.

In the four-mode system ($ n=4 $) in the linear configuration with
different damping and/or amplification rates in neighbor modes and
assuming $ \alpha_+ = 0 $ [$ \alpha_- = 0 $], $ \alpha_\pm =
\sqrt{ (3\pm\sqrt{5}) \zeta^2 /2 - \gamma_{-}^2)} = 0 $, we find a
single QHP with second-order ED and DD. The eigenvalues $
\Lambda_{1,2,3,4}^{M^{(4)}_{\rm l2}} $ [$
\Lambda_{5,6,7,8}^{M^{(4)}_{\rm l2}} $] in Eq.~(\ref{A1}) are
equal. The eigenvectors in Eq.~(\ref{A2}) fulfil: $\bm{
Y_{1}^{M^{(4)}_{\rm l2}} }= \bm{ Y_{3}^{M^{(4)}_{\rm l3}} } $ and
$\bm{ Y_{2}^{M^{(4)}_{\rm l2}} }= \bm{ Y_{4}^{M^{(4)}_{\rm l2}}}$
[$\bm{ Y_{5}^{M^{(4)}_{\rm l2}} }= \bm{ Y_{7}^{M^{(4)}_{\rm l2}} }
$ and $ \bm{ Y_{6}^{M^{(4)}_{\rm l2}} }= \bm{ Y_{8}^{M^{(4)}_{\rm
l2}}} $].

In the four-mode system ($ n=4 $) in the circular configuration
with equal damping and/or amplification rates of neighbor modes
and assuming $ \beta = \sqrt{\zeta^2 - \gamma_{-}^2} = 0 $, we
have two QHPs with second-order EDs and second-order DDs. The
eigenvalues and eigenvectors in this case behave analogous to
those given above for the four-mode system in the linear
configuration with equal damping and/or amplification rates of
neighbor modes.

\subsection{Five-mode bosonic systems}

In the five-mode system ($ n=5 $) in the linear configuration and
assuming $ \beta = \sqrt{ \zeta^2/4 - \gamma_{-}^2 } = 0 $, we
have two QHPs with second-order EDs and second-order DDs. The
eigenvalues $ \Lambda_{3,4,7,8}^{M^{(5)}_{\rm l}} $ and also the
eigenvalues $ \Lambda_{5,6,9,10}^{M^{(5)}_{\rm l}} $ in
Eq.~(\ref{A1}) are the same. The eigenvectors in Eq.~(\ref{A2})
fulfill the relations: $\bm{ Y_{3}^{M^{(5)}_{\rm l}} }= \bm{
Y_{7}^{M^{(5)}_{\rm l}} } $, $\bm{ Y_{4}^{M^{(5)}_{\rm l}} }= \bm{
Y_{8}^{M^{(5)}_{\rm l}} } $, $\bm{ Y_{5}^{M^{(5)}_{\rm l}} }= \bm{
Y_{9}^{M^{(5)}_{\rm l}} } $, and $\bm{ Y_{6}^{M^{(5)}_{\rm l}} }=
\bm{ Y_{10}^{M^{(5)}_{\rm l}} } $.

In the five-mode system ($ n=5 $) in the pyramid configuration and
assuming $ \beta_1 = \sqrt{ \zeta^2 - \gamma_{-}^2 } = 0 $, we
have two QEPs with second-order ED. The eigenvalues $
\Lambda_{3}^{M^{(5)}_{\rm p}} $ and $ \Lambda_{5}^{M^{(5)}_{\rm
p}} $ together with their accompanying eigenvectors $\bm{
Y_{3}^{M^{(5)}_{\rm p}}} = \bm{ Y_{5}^{M^{(5)}_{\rm p}}} $
coincide. The same is true for eigenvalues $
\Lambda_{4}^{M^{(5)}_{\rm p}} $ and $ \Lambda_{6}^{M^{(5)}_{\rm
p}} $ and eigenvectors $\bm{ Y_{4}^{M^{(5)}_{\rm p}}} $ and $ \bm{
Y_{6}^{M^{(5)}_{\rm p}}} $. Provided that $ \beta_2 = \sqrt{
5\zeta^2 - \gamma_{-}^2 } = 0 $, we reveal two QEPs with
second-order ED. The eigenvalues $ \Lambda_{7}^{M^{(5)}_{\rm p}} $
and $ \Lambda_{9}^{M^{(5)}_{\rm p}} $ together with their
accompanying eigenvectors $\bm{ Y_{7}^{M^{(5)}_{\rm p}}} $ and
$\bm{ Y_{9}^{M^{(5)}_{\rm p}}} $ are the same. Similarly, the
eigenvalues $ \Lambda_{8}^{M^{(5)}_{\rm p}} $ and $
\Lambda_{10}^{M^{(5)}_{\rm p}} $ and eigenvectors $\bm{
Y_{8}^{M^{(5)}_{\rm p}}} $ and $ \bm{ Y_{10}^{M^{(5)}_{\rm p}}} $
equal.

\section{QEPs and QHPs in first- and second-order field-operator-moments spaces:
Comparison}
\label{AppB}

Here, we present tables that summarize the QEPs and QHPs, along
with their degeneracies, as observed in the first- and
second-order FOM spaces for the bosonic systems analyzed in the
main text. This analysis follows the methodology developed in
Ref.~\cite{PerinaJr2022a}, which derives the genuine and induced
QEPs and QHPs --- together with their degeneracies --- from the
inherited QEPs and QHPs identified in the dynamics of first-order
FOMs.

The genuine QEPs can be viewed as analogues of the inherited QEPs
in the dynamical matrix of first-order FOMs, now extended to the
dynamical matrices of higher-order FOMs. However, a crucial
distinction arises due to the structure of higher-order FOMs,
which are constructed as products of a fixed number of field
operators, incorporating all combinations of annihilation and
creation operators. These combinations include terms that are
mutually related by commutation relations. Since such terms
contribute to the system dynamics only once, only one
representative term should be considered when analyzing spectral
degeneracies --- this yields the genuine QEPs.

In contrast, if all such terms are retained in the analysis,
without accounting for redundancy due to commutation relations,
the resulting degeneracies are referred to as induced QEPs. The
same distinction applies to QHPs, leading to the identification of
both genuine and induced QHPs~\cite{PerinaJr2022a}. Different
configurations of QEPs and QHPs emerge across different models,
highlighting the rich variety of spectral degeneracy structures
and their implications for diverse physical phenomena.

For the two-mode bosonic system with its dynamical matrix ${\bm
M}^{(2)}$ given in Eq.~(\ref{3}), we summarize the corresponding
degeneracies up to second-order FOMs in Tab.~\ref{tab1}. This
provides a direct comparison of the QEP and QHP degeneracies
observed in the dynamics of first- and second-order FOMs. Notably,
we identify genuine QEPs with a fourth-order ED in the
second-order FOM dynamics, in contrast to the inherited QEPs
exhibiting only second-order EDs in the first-order FOM dynamics.

It is instructive to compare the QEP and QHP degeneracies listed
in Tab.~\ref{tab1} with those observed in systems featuring larger
numbers of modes, as presented below. This comparison highlights
how the spectral degeneracy structures in first-order FOMs evolve
when extended to second-order FOMs. Notably, characteristic
features emerge that can be generalized to spectral degeneracies
of arbitrary-order FOMs, as discussed in
Ref.~\cite{PerinaJr2022a}. For instance, one can infer the maximal
EDs and maximal DDs occurring in the dynamics of FOMs of a given
order.
\begin{table*}[t]    
\begin{center}
\begin{tabular}{|p{0.4cm}|p{0.8cm}|p{2.4cm}|p{1.3cm}|p{1.3cm}|p{1cm}|p{1cm}|p{1cm}|p{1cm}|}
 \hline
  $ \bm{ \Lambda^{\rm i}_{j}} $ & $ \bm{ \Lambda^{\rm r}_{j}} $ & \multicolumn{2}{c|}{Moments} & Moment & \multicolumn{2}{p{2.5cm}|}{Genuine and induced QHPs} &  \multicolumn{2}{c|}{Genuine QHPs}\\
 \cline{6-9}
   &  &  \multicolumn{2}{c|}{} & deg. & Partial & Partial & Partial & Partial \\
  &  &  \multicolumn{2}{c|}{}  &  & QDP x & QDP x & QDP x & QDP x  \\
    &  &  \multicolumn{2}{c|}{}  &  & QEP deg. & QEP deg. & QEP deg. & QEP deg.\\
 \hline
 \hline
  $ \gamma_+ $ & $ \mp \beta $ & $ \langle \hat{b}_1 \rangle $, $ \langle \hat{b}_1^\dagger \rangle $ & $ \langle \bm{\hat{B}_1} \rangle $ & 1 & 1x2 & 2x2  & 1x2 & 2x2\\
 \cline{3-6}   \cline{8-8}
  &  & $ \langle \hat{b}_2 \rangle $, $ \langle \hat{b}_2^\dagger \rangle $ & $ \langle \bm{\hat{B}_2}  \rangle $ &  1 & 1x2 & & 1x2 &  \\
 \hline
  $ 2 \gamma_+ $ & $ \mp 2\beta $ & $ \langle \hat{b}_1 \hat{b}_2 \rangle $, $ \langle \hat{b}_1^\dagger \hat{b}_2^\dagger \rangle $ & $\langle \bm{\hat{B}_1\hat{B}_2} \rangle $ &  2 & 2x4 & 4x4 & 1x4 & 1x4 \\
      & $ \beta - \beta $  &  $ \langle \hat{b}_1^\dagger \hat{b}_2 \rangle $ &  & 2 & & & & + \\
      & $ \beta - \beta $  &  $ \langle \hat{b}_1 \hat{b}_2^\dagger \rangle $ & &  2 &  & & & \\
 \cline{2-6}   \cline{8-8}
  & $ \mp 2\beta $ & $ \langle \hat{b}_1^2 \rangle $, $ \langle \hat{b}_1^{\dagger 2}\rangle $ & $ \langle \bm{\hat{B}_1^2}  \rangle $ &  1 & 1x4 &  & 1x3 & 2x3\\
  & $ \beta - \beta $  & $ \langle \hat{b}_1^\dagger \hat{b}_1 \rangle $ & &  2 & & & & \\
 \cline{2-6}  \cline{8-8}
  & $ \mp 2\beta $ & $ \langle \hat{b}_2^2 \rangle $, $ \langle \hat{b}_2^{\dagger 2}\rangle $ & $ \langle \bm{\hat{B}_2^2}  \rangle $ & 1 & 1x4 &  & 1x3 &  \\
      & $ \beta - \beta $  &  $ \langle \hat{b}_2^\dagger \hat{b}_2 \rangle $ &  &  2 & &  & &\\
  \hline
\end{tabular} 
\end{center}
\vspace{-5mm}
 \caption{Real and imaginary parts of the complex eigenfrequencies $
  \bm{\Lambda_{j}^{\rm r}} - i
  \bm{\Lambda_{j}^{\rm i}} $ of the matrix $ \bm{M^{(2)}} $ given in Eq.~(\ref{3})
   for the two-mode bosonic system derived from the equations for
   the FOMs up to second order. The corresponding moments in the
  `diagonalized' field operators are written together with their degeneracies (deg.) coming from
  different positions of the field operators. The DDs of QHPs (partial DDs) derived from
  the indicated FOMs and the EDs of the constituting QEPs are given. Both genuine and induced QEPs and QHPs are considered.
   {The operator vectors $ \bm{\hat{B}_j} $ for
  $ j=1,2 $ are defined in the rows written for $ \bm{ \Lambda^{\rm i}_{j}} = \gamma_+ $
  devoted to the first-order FOMs, i.e. $ \bm{\hat{B}_j} \equiv [\hat{b}_j,\hat{b}_j^\dagger] $ for $ j=1,2$.
  Symbol $ \bm{\hat{B}_j \hat{B}_k} $, $ j,k=1,2 $, stands for the tensorial product
  that provides four terms explicitly written in the rows for $ \bm{ \Lambda^{\rm i}_{j}} = 2\gamma_+
  $; the terms derived from those explicitly written by using the commutation relations are omitted.}}
\label{tab1}
\end{table*}

Considering the linear three-mode bosonic system with its
dynamical matrix ${\bm M}^{(3)}$ given in Eq.~(\ref{21}) under the
condition given in Eq.~(\ref{22}), the QEPs and QHPs predicted in
the dynamics of first- and second-order FOMs are summarized in
Tab.~\ref{tab2}. As shown there, the inherited QEPs with a
third-order ED give rise to genuine QEPs with a ninth-order ED
when analyzing the dynamics of second-order FOMs.
\begin{table*}   
 \begin{center}
 \begin{tabular}{|p{0.4cm}|p{1.2cm}|p{2.4cm}|p{1.3cm}|p{1.3cm}|p{1cm}|p{1cm}|p{1cm}|p{1cm}|}
 \hline
  $ \bm{ \Lambda^{\rm i}_{j}} $ & $ \bm{ \Lambda^{\rm r}_{j}} $ & \multicolumn{2}{c|}{Moments} & Moment & \multicolumn{2}{p{2.5cm}|}{Genuine and induced QHPs} &  \multicolumn{2}{c|}{Genuine QHPs}\\
 \cline{6-9}
   &  &  \multicolumn{2}{c|}{} & deg. & Partial & Partial & Partial & Partial \\
  &  &  \multicolumn{2}{c|}{}  &  & QDP x & QDP x & QDP x & QDP x  \\
    &  &  \multicolumn{2}{c|}{}  &  & QEP deg. & QEP deg. & QEP deg. & QEP deg.\\
 \hline
  \hline
    $\gamma_+$ & 0, $\mp \beta$ & $\langle \hat{b}_1 \rangle$, $\langle \hat{b}_2 \rangle$, $\langle \hat{b}_2^\dagger \rangle$ & $ \langle \bm{\hat{B}_1} \rangle $ & 1 & 1x3 & 2x3 & 1x3 & 2x3  \\
    \cline{2-6}  \cline{8-8}
    & 0, $ \mp  \beta$ & $\langle \hat{b}_1^\dagger \rangle$,  $\langle \hat{b}_3 \rangle$, $\langle \hat{b}_3^\dagger \rangle$  & $ \langle \bm{\hat{B}_2}  \rangle $ & 1 & 1x3 & & 1x3 & \\
  \hline
   $2\gamma_+$ & $ 0$ & $\langle \hat{b}_1^2 \rangle$ & $ \langle \bm{\hat{B}_1^2}  \rangle $ & 1 & 1x9 & 4x9 & 1x6 & 2x6 \\
   & $\mp  \beta$ & $\langle \hat{b}_1 \hat{b}_2 \rangle$, $\langle \hat{b}_1 \hat{b}_2^\dagger  \rangle$   & & 2 &  &  &  & + \\
   & $ \mp 2\beta$ & $\langle \hat{b}_2^2 \rangle$, $\langle \hat{b}_2^{\dagger 2} \rangle$   & & 1 &  & &  & \\
   & $ \beta-\beta$ & $\langle \hat{b}_2^{\dagger } \hat{b}_2 \rangle$ & & 2 &  & &  & \\
   \cline{2-6}   \cline{8-8}
   & 0 &  $\langle \hat{b}_1^{\dagger } \hat{b}_1  \rangle$  & $\langle \bm{\hat{B}_2\hat{B}_1} \rangle $  & 2 & 2x9 &  & 1x9 & 1x9 \\
    & $\mp  \beta$ & $\langle \hat{b}_1^\dagger  \hat{b}_2   \rangle$, $\langle \hat{b}_1^\dagger   \hat{b}_2^\dagger   \rangle$   & & 2 &  & &  & \\
   & $\mp  \beta$ & $\langle \hat{b}_3 \hat{b}_1  \rangle$,  $\langle \hat{b}_3^\dagger  \hat{b}_1  \rangle$  & & 2 &  & &  & \\
   & $ \mp 2\beta$ & $\langle \hat{b}_3 \hat{b}_2 \rangle$, $\langle \hat{b}_3^\dagger \hat{b}_2^\dagger  \rangle$  & & 2 &  & &  & \\
   & $ \beta-\beta$ & $\langle \hat{b}_3  \hat{b}_2^\dagger \rangle$, $\langle \hat{b}_3^\dagger \hat{b}_2 \rangle$  & & 2 &  & &  & \\
  \cline{2-6}  \cline{8-8}
   & $ 0$ & $\langle \hat{b}_1^{\dagger 2} \rangle$  & $ \langle \bm{\hat{B}_2^2}  \rangle $  & 1 & 1x9 &  & 1x6 & \\
   & $\mp \beta$ & $\langle  \hat{b}_1^\dagger \hat{b}_3 \rangle$, $\langle \hat{b}_1^\dagger \hat{b}_3^\dagger \rangle$  & & 2 &  &  &  &\\
   & $ \mp 2\beta$ & $\langle \hat{b}_3^2 \rangle$, $\langle \hat{b}_3^{\dagger 3} \rangle$  & & 1 &  & &  & \\
   & $ \beta-\beta$ & $\langle \hat{b}_3^{\dagger } \hat{b}_3 \rangle$ & & 2 &  & &  & \\
  \hline
  \end{tabular}
 \end{center}
 \vspace{-5mm}
  \caption{Real and imaginary parts of the complex eigenfrequencies $
  \bm{\Lambda_{j}^{\rm r} } - i
  \bm{\Lambda_{j}^{\rm i} } $ of the matrix $ \bm{M^{(3)} } $, given in Eq.~(\ref{21}),
   for the linear three-mode bosonic system
   derived from the equations for the FOMs up to second order.  {We
   have $ \bm{\hat{B}_1} \equiv [\hat{b}_1,\hat{b}_2,\hat{b}_2^\dagger] $
   and $ \bm{\hat{B}_2} \equiv [\hat{b}_1^\dagger,\hat{b}_3,\hat{b}_3^\dagger] $
   and more details are given in the caption to Tab.~\ref{tab1}.}}
\label{tab2}
\end{table*}

In the linear four-mode bosonic system described by the dynamical
matrix ${\bm M}_{\rm l}^{(4)}$ given in Eq.~(\ref{28}), which
features equal damping and/or amplification rates for neighboring
modes [see Eq.~(\ref{29})], the QEPs and QHPs in the dynamics of
first- and second-order FOMs are summarized in Tab.~\ref{tab3}.
These results also apply to the circular four-mode bosonic system
with dynamical matrix ${\bm M}_{\rm c}^{(4)}$ shown in
Eq.~(\ref{38}), assuming equal damping and/or amplification rates
of neighboring modes [see Eq.~(\ref{39}]. According to
Tab.~\ref{tab3}, the inherited QEPs with doubled second-order ED
give rise to genuine QEPs exhibiting fourth-order ED and
sixth-order DD when analyzing second-order FOM dynamics.
\begin{table*}   
 \begin{center}
  \begin{tabular}{|p{0.4cm}|p{1.6cm}|p{2.2cm}|p{1.3cm}|p{1.3cm}|p{1cm}|p{1cm}|p{1cm}|p{1cm}|}
 \hline
  $ \bm{ \Lambda^{\rm i}_{j}} $ & $ \bm{ \Lambda^{\rm r}_{j}} $ & \multicolumn{2}{c|}{Moments} & Moment & \multicolumn{2}{p{2.5cm}|}{Genuine and induced QHPs} &  \multicolumn{2}{c|}{Genuine QHPs}\\
 \cline{6-9}
   &  &  \multicolumn{2}{c|}{} & deg. & Partial & Partial & Partial & Partial \\
  &  &  \multicolumn{2}{c|}{}  &  & QDP x & QDP x & QDP x & QDP x  \\
    &  &  \multicolumn{2}{c|}{}  &  & QEP deg. & QEP deg. & QEP deg. & QEP deg.\\
 \hline
  \hline
  $\gamma_+$ & $ \alpha_{\pm}$ & $\langle \hat{b}_1 \rangle$, $\langle \hat{b}_3 \rangle$ & $ \langle \bm{\hat{B}_1} \rangle $ & 1 & 2x2 & 4x2 & 2x2 & 4x2 \\
  \cline{3-5}
  &   & $\langle \hat{b}_4 \rangle$, $\langle \hat{b}_2 \rangle$ & $ \langle \bm{\hat{B}_2}  \rangle $ & 1 &  & &  & \\
  \cline{2-6} \cline{8-8}
   & $ -\alpha_{\pm}$ &  $\langle \hat{b}_1^\dagger \rangle$, $\langle \hat{b}_3 ^\dagger\rangle$  & $\langle \bm{\hat{B}_3} \rangle $ & 1 & 2x2 &  & 2x2 & \\
  \cline{3-5}
  &   & $\langle \hat{b}_4^\dagger \rangle$, $\langle \hat{b}_2^\dagger \rangle$  & $\langle \bm{\hat{B}_4} \rangle $ & 1 & & & & \\
  \hline
  $2\gamma_+$ & $ 2\alpha_{\pm}$ & $\langle \hat{b}_1^2 \rangle$, $\langle \hat{b}_3^2 \rangle$ & $ \langle \bm{\hat{B}_1^2}  \rangle $ & 1 & 1x4 & 16x4 & 1x3 & 4x3  \\
   & $ \alpha_{-}+\alpha_{+}$ & $\langle \hat{b}_1 \hat{b}_3 \rangle$ & & 2 &  &  & & + \\
  \cline{2-6} \cline{8-8}
   & $ 2\alpha_{\mp}$ & $\langle \hat{b}_2^2 \rangle$, $\langle \hat{b}_4^2 \rangle$ & $\langle \bm{\hat{B}_2^2} \rangle $ & 1 & 1x4 & & 1x3 & \\
  & $ \alpha_{-}+\alpha_{+}$ & $\langle \hat{b}_2 \hat{b}_4 \rangle$ &  & 2 &  & &  & \\
  \cline{2-6} \cline{8-8}
   &  $- 2\alpha_{\pm}$ & $\langle \hat{b}_1^{\dagger 2} \rangle$, $\langle \hat{b}_3^{\dagger 2} \rangle$   & $ \langle \bm{\hat{B}_3^2}  \rangle $ & 1 & 1x4 &  & 1x3 &\\
   & $ -\alpha_{-}-\alpha_{+}$ & $\langle \hat{b}_1^{\dagger } \hat{b}_3^{\dagger } \rangle$ &  & 2 &  &  &  & \\
 \cline{2-6} \cline{8-8}
  &  $ -2\alpha_{\mp}$ & $\langle \hat{b}_2^{\dagger 2} \rangle$, $\langle \hat{b}_4^{\dagger 2} \rangle$   & $\langle \bm{\hat{B}_4^2} \rangle $ & 1 & 1x4 &  & 1x3 &\\
  & $ -\alpha_{-}-\alpha_{+}$ & $\langle \hat{b}_2^{\dagger } \hat{b}_4^{\dagger } \rangle$ &  & 2 &  & &  & \\
  \cline{2-6} \cline{8-8}
  &  $ -\alpha_{\pm}+\alpha_{\pm}$ & $\langle \hat{b}_1^{\dagger} \hat{b}_1 \rangle$, $\langle \hat{b}_3^{\dagger} \hat{b}_3 \rangle$  & $\langle \bm{\hat{B}_3\hat{B}_1} \rangle $ & 2 & 2x4 & & 1x4 & 6x4\\
  & $ \pm\alpha_{-}\mp\alpha_{+}$ & $\langle  \hat{b}_1^\dagger \hat{b}_3 \rangle$, $\langle \hat{b}_1 \hat{b}_3^\dagger \rangle$  &  & 2 &  &  &  &\\
  \cline{2-6} \cline{8-8}
  & $ -\alpha_{\mp}+\alpha_{\mp}$ & $\langle \hat{b}_2^\dagger \hat{b}_2 \rangle$, $\langle \hat{b}_4^\dagger \hat{b}_4 \rangle$  & $\langle \bm{\hat{B}_4\hat{B}_2} \rangle $  & 2 & 2x4 &   & 1x4 & \\
  & $\mp\alpha_{-}\pm\alpha_{+}$ & $\langle  \hat{b}_2^\dagger \hat{b}_4 \rangle$, $\langle \hat{b}_4^\dagger \hat{b}_2 \rangle$  &  & 2 &  & &  & \\
  \cline{2-6} \cline{8-8}
  & $ \alpha_{+}+\alpha_{-}$ & $\langle \hat{b}_1 \hat{b}_2 \rangle$, $\langle \hat{b}_3 \hat{b}_4 \rangle$  & $\langle \bm{\hat{B}_1\hat{B}_2} \rangle $ & 2 & 2x4 & & 1x4 & \\
  & $ 2\alpha_{\pm}$ & $\langle  \hat{b}_1 \hat{b}_4 \rangle$, $\langle \hat{b}_3 \hat{b}_2 \rangle$  &  & 2 &  & &  & \\
  \cline{2-6} \cline{8-8}
  & $ \alpha_{\mp}-\alpha_{\pm}$ & $\langle \hat{b}_1^\dagger \hat{b}_2 \rangle$, $\langle \hat{b}_3^\dagger \hat{b}_4 \rangle$  & $\langle \bm{\hat{B}_3\hat{B}_2} \rangle $ & 2 & 2x4 & & 1x4 & \\
  & $ \alpha_{\pm}-\alpha_{\pm}$ & $\langle  \hat{b}_1^\dagger \hat{b}_4 \rangle$, $\langle \hat{b}_3^\dagger \hat{b}_2 \rangle$  &  & 2 &  & &  & \\
  \cline{2-6} \cline{8-8}
  & $ \alpha_{\pm}-\alpha_{\mp}$ & $\langle \hat{b}_2^\dagger \hat{b}_1 \rangle$, $\langle \hat{b}_4^\dagger \hat{b}_3 \rangle$  & $ \langle \bm{\hat{B}_4\hat{B}_1}  \rangle$ & 2 & 2x4 & & 1x4 & \\
  & $ \alpha_{\pm}-\alpha_{\pm}$ & $\langle  \hat{b}_4^\dagger \hat{b}_1 \rangle$, $\langle \hat{b}_2^\dagger \hat{b}_3 \rangle$  &  & 2 &  & &  &  \\
  \cline{2-6} \cline{8-8}
  & $ -\alpha_{+}-\alpha_{-}$ & $\langle \hat{b}_1^\dagger \hat{b}_2^\dagger \rangle$, $\langle \hat{b}_3^\dagger \hat{b}_4^\dagger \rangle$  & $\langle \bm{\hat{B}_3\hat{B}_4} \rangle $ & 2 & 2x4 & & 1x4 & \\
  & $- 2\alpha_{\pm}$ & $\langle  \hat{b}_1^\dagger \hat{b}_4^\dagger \rangle$, $\langle \hat{b}_3^\dagger \hat{b}_2^\dagger \rangle$  &  & 2 &  &  &  & \\
 \hline
 \end{tabular}
 \end{center}
 \vspace{-5mm}
  \caption{Real and imaginary parts of the complex eigenfrequencies $
  \bm{\Lambda_{j}^{\rm r} } - i
  \bm{\Lambda_{j}^{\rm i} } $ of the matrix $ \bm{M^{(4)}_{\rm l1}} $, given in Eq.~(\ref{28}) with Eq.~(\ref{29}),
   for the linear four-mode bosonic system
   with equal damping and/or amplification rates for neighbor modes derived from the equations for
   the FOMs up to second order. We note that $ {\alpha}_{\pm}\left( \zeta\right)={\alpha}_{\mp}\left(-\zeta\right)$ is used here.
    {We have $ \bm{\hat{B}_1} \equiv [\hat{b}_1,\hat{b}_3] $, $ \bm{\hat{B}_2} \equiv [\hat{b}_2,\hat{b}_4] $,
   $ \bm{\hat{B}_3} = \bm{\hat{B}_1}^\dagger $, $ \bm{\hat{B}_4} = \bm{\hat{B}_2}^\dagger
   $. More details are given in the caption to Tab.~\ref{tab1}.}}
\label{tab3}
\end{table*}

On the other hand, for the linear four-mode bosonic system
described by the dynamical matrix ${\bm M}_{\rm l}^{(4)}$ given in
Eq.~(\ref{28}), with varying damping and/or amplification rates of
neighboring modes [see Eq.~(\ref{33})], the QEPs and QHPs observed
in the dynamics of first- and second-order FOMs are summarized in
Tab.~\ref{tab4}. The spectral structure of first-order FOMs,
featuring two inherited QEPs with second-order EDs, evolves into a
more complex second-order FOM structure that includes genuine QEPs
with second-, third-, and fourth-order EDs. Notably, the genuine
QHPs with second-order ED exhibit a sixteenth-order DD.
\begin{table*}   
 \begin{center}
   \begin{tabular}{|p{0.4cm}|p{1.6cm}|p{2.2cm}|p{1.3cm}|p{1.3cm}|p{1cm}|p{1cm}|p{1cm}|p{1cm}|}
 \hline
  $ \bm{ \Lambda^{\rm i}_{j}} $ & $ \bm{ \Lambda^{\rm r}_{j}} $ & \multicolumn{2}{c|}{Moments} & Moment & \multicolumn{2}{p{2.5cm}|}{Genuine and induced QHPs} &  \multicolumn{2}{c|}{Genuine QHPs}\\
 \cline{6-9}
   &  &  \multicolumn{2}{c|}{} & deg. & Partial & Partial & Partial & Partial \\
  &  &  \multicolumn{2}{c|}{}  &  & QDP x & QDP x & QDP x & QDP x  \\
    &  &  \multicolumn{2}{c|}{}  &  & QEP deg. & QEP deg. & QEP deg. & QEP deg.\\
 \hline
   \hline
  $\gamma_+$ & $\pm \alpha_{+}$ & $\langle \hat{b}_1 \rangle$, $\langle \hat{b}_1^\dagger  \rangle$ & $ \langle \bm{\hat{B}_1} \rangle $ & 1 & 1x2 & 2x2  & 1x2 & 2x2 \\
  \cline{3-6}  \cline{8-8}
   &  & $\langle \hat{b}_2\rangle$,  $\langle \hat{b}_2^\dagger \rangle$  & $ \langle \bm{\hat{B}_2}  \rangle $ & 1 & 1x2 &  & 1x2 & \\
  \cline{2-9}
   & $ \alpha_{-}$ & $\langle \hat{b}_3 \rangle$  & $\langle \bm{\hat{B}_3} \rangle $ & 1 & 2x1 & 2x1 & 2x1 & 2x1 \\
  \cline{3-5}
  &   &  $\langle \hat{b}_4 \rangle$ & $\langle \bm{\hat{B}_4} \rangle $ & 1 &  & &  & \\
  \cline{2-9}
  & $-\alpha_{-}$ & $\langle \hat{b}_3^\dagger \rangle$  & $\langle \bm{\hat{B}_5} \rangle $ & 1 & 2x1 & 2x1 & 2x1 & 2x1 \\
  \cline{3-5}
  &   &   $\langle \hat{b}_4^\dagger \rangle$  & $\langle \bm{\hat{B}_6} \rangle $ & 1 &  & &  &  \\
 \hline
  $2\gamma_+$ & $\pm2 \alpha_{+}$ & $\langle \hat{b}_1^2 \rangle$, $\langle \hat{b}_1^{\dagger 2}\rangle$ & $ \langle \bm{\hat{B}_1^2}  \rangle $ & 1 & 1x4 & 4x4 & 1x3 & 2x3 \\
   & $ \alpha_{+}-\alpha_{+}$ &  $\langle \hat{b}_1^{\dagger } \hat{b}_1  \rangle$ & & 2 &  & &  & + \\
  \cline{2-6} \cline{8-8}
  &  $\pm 2 \alpha_{+}$ & $\langle \hat{b}_2^{2} \rangle$, $\langle \hat{b}_2^{\dagger 2} \rangle$   & $ \langle \bm{\hat{B}_2^2}  \rangle $ & 1 & 1x4 & & 1x3 & \\
  & $ \alpha_{+}-\alpha_{+}$ & $\langle \hat{b}_2^{\dagger } \hat{b}_2 \rangle$ & & 2 &  & &  & \\
  \cline{2-6} \cline{8-8}
   & $\pm \alpha_{+}$ &  $\langle \hat{b}_1  \hat{b}_2  \rangle$, $\langle \hat{b}_1^{\dagger } \hat{b}_2^{\dagger }  \rangle$  & $\langle \bm{\hat{B}_1\hat{B}_2} \rangle $ & 2 & 2x4 & & 1x4 & 1x4 \\
    & $ \alpha_{+}-\alpha_{+}$ & $\langle \hat{b}_1^\dagger  \hat{b}_2   \rangle$, $\langle  \hat{b}_1 \hat{b}_2^\dagger  \rangle$ &  & 2 &  & &  & \\
  \cline{2-9}
  & $ \alpha_{-}\pm\alpha_{+}$ & $\langle \hat{b}_1 \hat{b}_3 \rangle$, $\langle  \hat{b}_1^\dagger \hat{b}_3 \rangle$  & $\langle \bm{\hat{B}_1\hat{B}_3} \rangle $  & 2 & 2x2 & 16x2 & 2x2 & 16x2 \\
   \cline{3-6} \cline{8-8}
  &   & $\langle \hat{b}_1 \hat{b}_4 \rangle$, $\langle  \hat{b}_1^\dagger \hat{b}_4 \rangle$  & $\langle \bm{\hat{B}_1\hat{B}_4} \rangle $ & 2 &  2x2 & &  2x2 &\\
  \cline{2-6} \cline{8-8}
  & $ \alpha_{-}\pm\alpha_{+}$ & $\langle \hat{b}_2 \hat{b}_3 \rangle$, $\langle \hat{b}_2^\dagger \hat{b}_3 \rangle$ & $\langle \bm{\hat{B}_2\hat{B}_3} \rangle $  & 2 & 2x2 & & 2x2 & \\
  \cline{3-6} \cline{8-8}
  &   & $\langle \hat{b}_2 \hat{b}_4 \rangle$, $\langle \hat{b}_2^\dagger \hat{b}_4 \rangle$ & $\langle \bm{\hat{B}_2\hat{B}_4} \rangle $  & 2 & 2x2 & & 2x2 &\\
  \cline{2-6} \cline{8-8}
  & $ -\alpha_{-}\pm\alpha_{+}$ & $\langle \hat{b}_1 \hat{b}_3^\dagger \rangle$, $\langle \hat{b}_1^\dagger \hat{b}_3^\dagger \rangle$ & $\langle \bm{\hat{B}_1\hat{B}_5} \rangle $  & 2 & 2x2 & & 2x2 & \\
   \cline{3-6} \cline{8-8}
  &   & $\langle  \hat{b}_1 \hat{b}_4^\dagger \rangle$, $\langle \hat{b}_1^\dagger \hat{b}_4^\dagger \rangle$ & $\langle \bm{\hat{B}_1\hat{B}_6} \rangle $  & 2 & 2x2 & & 2x2 &\\
  \cline{2-6}  \cline{8-8}
  & $- \alpha_{-}\pm\alpha_{+}$ & $\langle \hat{b}_2 \hat{b}_3^\dagger \rangle$, $\langle \hat{b}_2^\dagger \hat{b}_3^\dagger \rangle$ & $\langle \bm{\hat{B}_2\hat{B}_5} \rangle $  & 2 & 2x2 & & 2x2 & \\
  \cline{3-6}  \cline{8-8}
  &  & $\langle  \hat{b}_2 \hat{b}_4^\dagger \rangle$, $\langle \hat{b}_2^\dagger \hat{b}_4^\dagger \rangle$ & $\langle \bm{\hat{B}_2\hat{B}_6} \rangle $  & 2 & 2x2  & & 2x2  &\\
 \cline{2-9}
  & $2 \alpha_{-}$ & $\langle \hat{b}_3^2 \rangle$, $\langle \hat{b}_4^2 \rangle$  & $\langle \bm{\hat{B}_3^2} \rangle $, $\langle \bm{\hat{B}_4^2} \rangle $  & 1 & 4x1 & 4x1 & 3x1 & 3x1 \\
 \cline{3-4}
  &  & $\langle \hat{b}_3 \hat{b}_4 \rangle$ & $\langle  \bm{\hat{B}_3\hat{B}_4}  \rangle$ & 2 &  & &  & \\
 \cline{2-9}
  & $-2 \alpha_{-}$ & $\langle \hat{b}_3^{\dagger 2} \rangle$, $\langle \hat{b}_4^{\dagger 2} \rangle$ & $\langle \bm{\hat{B}_5^2} \rangle $, $\langle \bm{\hat{B}_6^2} \rangle $   & 1 & 4x1 & 4x1 & 3x1 & 3x1 \\
 \cline{3-4}
  &   & $\langle \hat{b}_3^{\dagger } \hat{b}_4^{\dagger } \rangle$ & $ \langle \bm{\hat{B}_5\hat{B}_6}  \rangle$ & 2 &  &  &  &  \\
 \cline{2-9}
  & $ \alpha_{-}-\alpha_{-}$ & $\langle \hat{b}_3^{\dagger } \hat{b}_3 \rangle$, $\langle \hat{b}_4^{\dagger } \hat{b}_4  \rangle$ & $\langle  \bm{\hat{B}_5\hat{B}_3}  \rangle$, $\langle  \bm{\hat{B}_6\hat{B}_4} \rangle $ & 2 & 8x1 &  8x1 & 4x1 &  4x1 \\
 \cline{3-4}
  &  & $\langle \hat{b}_3^{\dagger } \hat{b}_4 \rangle$, $\langle \hat{b}_3  \hat{b}_4^{\dagger } \rangle$ & $\langle  \bm{\hat{B}_5\hat{B}_4} \rangle $, $\langle \bm{\hat{B}_3\hat{B}_6} \rangle $ & 2 &  &   &  &   \\
 \hline
 \end{tabular}
 \end{center}
 \vspace{-5mm}
  \caption{Real and imaginary parts of the complex eigenfrequencies $
  \bm{\Lambda_{j}^{\rm r} } - i
  \bm{\Lambda_{j}^{\rm i} } $ of the matrix $ \bm{M^{(4)}_{\rm l2} } $, given in Eq.~(\ref{28}) with Eq.~(\ref{33}) for the linear
   four-mode bosonic system
   with different damping and/or amplification rates of neighbor modes,
   as derived from the equations for the FOMs up to second order assuming $ \alpha_+ = 0 $.
    {We have $ \bm{\hat{B}_j} \equiv [\hat{b}_j,\hat{b}_j^\dagger] $ for $ j=1,2 $, $ \bm{\hat{B}_j} \equiv [\hat{b}_j] $
   for $ j=3,4 $, $ \bm{\hat{B}_5} = \bm{\hat{B}_3}^\dagger $, $ \bm{\hat{B}_6} =
   \bm{\hat{B}_4}^\dagger $ and more details are given in the caption to Tab.~\ref{tab1}.}}
\label{tab4}
\end{table*}

In the linear five-mode bosonic system described by the dynamical
matrix ${\bm M}_{\rm l}^{(5)}$ given in Eq.~(\ref{44}), with
damping and/or amplification rates satisfying the conditions in
Eqs.~(\ref{45}) and (\ref{46}), the QEPs and QHPs observed in the
dynamics of first- and second-order FOMs are summarized in
Tab.~\ref{tab5}. Notably, two inherited QEPs with second-order EDs
give rise to genuine QEPs exhibiting second-, third-, and
fourth-order EDs in various configurations within the spectral
structure of second-order FOMs.
\begin{table*}   
 \begin{center}
   \begin{tabular}{|p{0.4cm}|p{1.6cm}|p{2.2cm}|p{1.9cm}|p{1.3cm}|p{0.93cm}|p{0.93cm}|p{0.93cm}|p{0.93cm}|}
 \hline
  $ \bm{ \Lambda^{\rm i}_{j}} $ & $ \bm{ \Lambda^{\rm r}_{j}} $ & \multicolumn{2}{c|}{Moments} & Moment & \multicolumn{2}{p{2.5cm}|}{Genuine and induced QHPs} &  \multicolumn{2}{c|}{Genuine QHPs}\\
 \cline{6-9}
   &  &  \multicolumn{2}{c|}{} & deg. & Partial & Partial & Partial & Partial \\
  &  &  \multicolumn{2}{c|}{}  &  & QDP x & QDP x & QDP x & QDP x  \\
    &  &  \multicolumn{2}{c|}{}  &  & QEP deg. & QEP deg. & QEP deg. & QEP deg.\\
 \hline
  \hline
  $\gamma_+$ & $ \alpha_{\mp}$ & $\langle \hat{b}_2 \rangle$, $\langle \hat{b}_4 \rangle$ & $ \langle \bm{\hat{B}_1} \rangle $ & 1 & 2x2 & 4x2 & 2x2 & 4x2 \\
  \cline{3-5}
   &  & $\langle \hat{b}_3 \rangle$, $\langle \hat{b}_5 \rangle$ & $\langle \bm{\hat{B}_2} \rangle $ & 1 &  & & & \\
 \cline{2-6} \cline{8-8}
  &  $ -\alpha_{\mp}$ & $\langle \hat{b}_2^\dagger \rangle$, $\langle \hat{b}_4 ^\dagger\rangle$ & $ \langle \bm{\hat{B}_3}  \rangle $ & 1 & 2x2 & & 2x2 & \\
  \cline{3-5}
  &   & $\langle \hat{b}_3^\dagger \rangle$,  $\langle \hat{b}_5^\dagger \rangle$ & $\langle \bm{\hat{B}_4} \rangle $  & 1 & &  & & \\
 \cline{2-9}
  & 0 & $\langle \hat{b}_1 \rangle$ & $\langle \bm{\hat{B}_5} \rangle $ & 1 & 2x1 & 2x1 & 2x1 & 2x1 \\
 \cline{3-5}
  &  & $\langle \hat{b}_1^\dagger \rangle$ & $\langle \bm{\hat{B}_6} \rangle $ & 1 &  &   &  &\\
 \hline
 $2\gamma_+$ & $ 2\alpha_{\mp}$ & $\langle \hat{b}_2^2 \rangle$, $\langle \hat{b}_4^2 \rangle$ & $ \langle \bm{\hat{B}_1^2}  \rangle $ & 1 & 1x4 & 16x4 & 1x3 & 4x3 \\
 & $ \alpha_{-}+\alpha_{+}$ & $\langle \hat{b}_2 \hat{b}_4 \rangle$ & & 2 &  & &  & + \\
 \cline{2-6}  \cline{8-8}
  & $ 2\alpha_{\mp}$ & $\langle \hat{b}_3^2 \rangle$, $\langle \hat{b}_5^2 \rangle$ & $\langle \bm{\hat{B}_2^2} \rangle $ & 1 & 1x4 & & 1x3 & \\
  & $ \alpha_{-}+\alpha_{+}$ & $\langle \hat{b}_3 \hat{b}_5 \rangle$ & & 2 &  & &  & \\
  \cline{2-6}  \cline{8-8}
  &  $- 2\alpha_{\mp}$ & $\langle \hat{b}_2^{\dagger 2} \rangle$, $\langle \hat{b}_4^{\dagger 2} \rangle$  & $ \langle \bm{\hat{B}_3^2}  \rangle $  & 1 & 1x4 & & 1x3 & \\
  & $ -\alpha_{-}-\alpha_{+}$ & $\langle \hat{b}_2^{\dagger } \hat{b}_4^{\dagger } \rangle$ & & 2 &  &  &  &\\
  \cline{2-6}  \cline{8-8}
  &  $ -2\alpha_{\mp}$ & $\langle \hat{b}_3^{\dagger 2} \rangle$, $\langle \hat{b}_5^{\dagger 2} \rangle$  & $\langle \bm{\hat{B}_4^2} \rangle $  & 1 & 1x4 & & 1x3 & \\
  & $ -\alpha_{-}-\alpha_{+}$ & $\langle \hat{b}_3^{\dagger } \hat{b}_5^{\dagger } \rangle$ & & 2 &  &  &  &\\
  \cline{2-6}  \cline{8-8}
  & $ 2\alpha_{\mp}$ & $\langle \hat{b}_2 \hat{b}_3 \rangle$, $\langle \hat{b}_4 \hat{b}_5 \rangle$ & $\langle \bm{\hat{B}_1\hat{B}_2} \rangle $    & 2 & 2x4 & & 1x4 & \\
  & $ \alpha_{+}+\alpha_{-}$  & $\langle  \hat{b}_2 \hat{b}_5 \rangle$, $\langle \hat{b}_4 \hat{b}_3 \rangle$  & & 2 &  & &  & \\
  \cline{2-6}  \cline{8-8}
  &  $ -\alpha_{\mp}+\alpha_{\mp}$ & $\langle \hat{b}_2^{\dagger} \hat{b}_2 \rangle$, $\langle \hat{b}_4^{\dagger} \hat{b}_4 \rangle$ & $\langle \bm{\hat{B}_3\hat{B}_1} \rangle $  & 2 & 2x4 &  & 1x4 & 6x4 \\
  & $ -\alpha_{\mp}+\alpha_{\pm}$ & $\langle  \hat{b}_2^\dagger \hat{b}_4 \rangle$, $\langle \hat{b}_2 \hat{b}_4^\dagger \rangle$ & & 2 &  & &  & \\
  \cline{2-6}  \cline{8-8}
  & $ \alpha_{\mp}-\alpha_{\mp}$ & $\langle \hat{b}_3^\dagger \hat{b}_2 \rangle$, $\langle \hat{b}_5^\dagger \hat{b}_4 \rangle$ & $\langle \bm{\hat{B}_4\hat{B}_1} \rangle $   & 2 & 2x4 &  & 1x4 & \\
  & $ \alpha_{\pm}-\alpha_{\mp}$ & $\langle  \hat{b}_5^\dagger \hat{b}_2 \rangle$, $\langle \hat{b}_3^\dagger \hat{b}_4 \rangle$ & & 2 &  & &  &\\
  \cline{2-6} \cline{8-8}
  & $ \alpha_{\mp}-\alpha_{\mp}$ & $\langle \hat{b}_2^\dagger \hat{b}_3 \rangle$, $\langle \hat{b}_4^\dagger \hat{b}_5 \rangle$ & $\langle \bm{\hat{B}_3\hat{B}_2} \rangle $   & 2 & 2x4 &  & 1x4 & \\
  & $ \alpha_{\pm}-\alpha_{\mp}$ & $\langle  \hat{b}_2^\dagger \hat{b}_5 \rangle$, $\langle \hat{b}_4^\dagger \hat{b}_3 \rangle$ & & 2 &  & &  & \\
  \cline{2-6}  \cline{8-8}
  & $ -\alpha_{\mp}+\alpha_{\mp}$ & $\langle \hat{b}_3^\dagger \hat{b}_3 \rangle$, $\langle \hat{b}_5^\dagger \hat{b}_5 \rangle$ & $\langle \bm{\hat{B}_4\hat{B}_2} \rangle $   & 2 & 2x4 & & 1x4 & \\
  & $ -\alpha_{\mp}+\alpha_{\pm}$ & $\langle  \hat{b}_3^\dagger \hat{b}_5 \rangle$, $\langle \hat{b}_5^\dagger \hat{b}_3 \rangle$ & & 2 &  & &  & \\
  \cline{2-6}  \cline{8-8}
  & $- 2\alpha_{\mp}$ & $\langle \hat{b}_2^\dagger \hat{b}_3^\dagger \rangle$, $\langle \hat{b}_4^\dagger \hat{b}_5^\dagger \rangle$ & $\langle \bm{\hat{B}_3\hat{B}_4} \rangle $ & 2 & 2x4 & & 1x4 & \\
  & $ -\alpha_{+}-\alpha_{-}$  & $\langle  \hat{b}_2^\dagger \hat{b}_5^\dagger \rangle$, $\langle \hat{b}_4^\dagger \hat{b}_3^\dagger \rangle$ & & 2 &  & &  &\\
  \cline{2-9}
  & $ \alpha_{\mp}$ & $\langle \hat{b}_2 \hat{b}_1 \rangle$, $\langle \hat{b}_4 \hat{b}_1\rangle$ & $\langle \bm{\hat{B}_1\hat{B}_5} \rangle $ & 2 & 8x2 & 16x2  & 4x2 & 8x2 \\
  \cline{3-4}
  &  & $\langle \hat{b}_2 \hat{b}_1^\dagger \rangle$ $\langle \hat{b}_4 \hat{b}_1^\dagger \rangle$ & $\langle \bm{\hat{B}_1\hat{B}_6} \rangle $  & 2 &  & &  & \\
  \cline{3-4}
  &  & $\langle \hat{b}_3 \hat{b}_1 \rangle$, $\langle \hat{b}_5 \hat{b}_1\rangle$ & $\langle \bm{\hat{B}_2\hat{B}_5} \rangle $   & 2 &  & &  & \\
  \cline{3-4}
  &  & $\langle \hat{b}_3 \hat{b}_1^\dagger \rangle$ $\langle \hat{b}_5 \hat{b}_1^\dagger \rangle$ & $\langle \bm{\hat{B}_2\hat{B}_6} \rangle $  & 2 &  & &  & \\
  \cline{2-6}  \cline{8-8}
  & $- \alpha_{\mp}$ & $\langle \hat{b}_2^\dagger \hat{b}_4 \rangle$, $\langle \hat{b}_4^\dagger \hat{b}_1\rangle$ & $\langle \bm{\hat{B}_3\hat{B}_5} \rangle $   & 2 & 8x2 & & 4x2 &  \\
  \cline{3-4}
  &  & $\langle \hat{b}_2^\dagger \hat{b}_1^\dagger \rangle$ $\langle \hat{b}_4^\dagger \hat{b}_1^\dagger \rangle$ & $\langle \bm{\hat{B}_3\hat{B}_6} \rangle $  & 2 &  & &  & \\
  \cline{3-4}
   &  & $\langle \hat{b}_3^\dagger \hat{b}_1 \rangle$, $\langle \hat{b}_5^\dagger \hat{b}_1\rangle$ & $\langle \bm{\hat{B}_4\hat{B}_5} \rangle $   & 2 &  & &  &  \\
  \cline{3-4}
  &  & $\langle \hat{b}_3^\dagger \hat{b}_1^\dagger \rangle$ $\langle \hat{b}_5^\dagger \hat{b}_1^\dagger \rangle$ & $\langle \bm{\hat{B}_4\hat{B}_6} \rangle $  & 2 &  & &  & \\
  \cline{2-9}
  & 0 & $\langle \hat{b}_1^2 \rangle$, $\langle \hat{b}_1^{\dagger 2} \rangle$ & $\langle \bm{\hat{B}_5^2} \rangle $, $\langle \bm{\hat{B}_6^2} \rangle $  & 1 & 4x1 & 4x1 & 3x1 & 3x1 \\
  \cline{3-4}
  & & $\langle \hat{b}_1^\dagger \hat{b}_1 \rangle$  & $\langle \bm{\hat{B}_6\hat{B}_5} \rangle $  & 2 & & & &  \\
  \hline
 \end{tabular}
 \end{center}
 \vspace{-5mm}
  \caption{Real and imaginary parts of the complex eigenfrequencies $
  \bm{\Lambda_{j}^{\rm r} } - i
  \bm{\Lambda_{j}^{\rm i} } $ of the matrix $ \bm{M^{(5)}_{\rm l}} $, given in Eq.~(\ref{44}),
   for the five-mode linear bosonic system
   derived from the equations for the FOMs up to second order.
   {We have $ \bm{\hat{B}_1} \equiv [\hat{b}_2,\hat{b}_4] $, $ \bm{\hat{B}_2} \equiv [\hat{b}_3,\hat{b}_5] $,
   $ \bm{\hat{B}_3} = \bm{\hat{B}_1}^\dagger $, $ \bm{\hat{B}_4} =
   \bm{\hat{B}_2}^\dagger $, $ \bm{\hat{B}_5} \equiv
   [\hat{b}_1] $, $ \bm{\hat{B}_6} = \bm{\hat{B}_5}^\dagger $,
   and more details are given in the caption to Tab.~\ref{tab1}.}}
\label{tab5}
\end{table*}

\begin{table*}   
 \hspace{-15mm}
\footnotesize
\begin{tabular}{|c|c|c|c|c|c|c|c|c|}
 \hline
  $ \bm{ \Lambda^{\rm i}_{j}} $ & $ \bm{ \Lambda^{\rm r}_{j}} $ & \multicolumn{2}{c|}{Moments} & Moment & \multicolumn{2}{c|}{Genuine and induced QHPs} &  \multicolumn{2}{c|}{Genuine QHPs}\\
 \cline{6-9}
   &  &  \multicolumn{2}{c|}{} & deg. & Partial & Partial & Partial & Partial \\
  &  &  \multicolumn{2}{c|}{}  &  & QDP x & QDP x & QDP x & QDP x  \\
    &  &  \multicolumn{2}{c|}{}  &  & QEP deg. & QEP deg. & QEP deg. & QEP deg.\\
 \hline
  \hline
  $\gamma_+$ & $ -\zeta\mp\beta_{2}$ & $\langle \hat{b}_4 \rangle$, $\langle \hat{b}_5^\dagger \rangle$ & $ \langle \bm{\hat{B}_1} \rangle $ & 1 & 1x2 & 1x2 & 1x2 & 1x2 \\
  \cline{2-9}
  &  $ \zeta\pm\beta_{2}$ & $\langle \hat{b}_4^\dagger \rangle$, $\langle \hat{b}_5 \rangle$ & $ \langle \bm{\hat{B}_2}  \rangle $ & 1 & 1x2 & 1x2 & 1x2 & 1x2 \\
 \cline{2-9}
  & $ \zeta-\beta_{1}$ & $\langle \hat{b}_2 \rangle$ & $\langle \bm{\hat{B}_3} \rangle $ & 1 & 1x1 & 1x1 & 1x1 & 1x1 \\
 \cline{2-9}
  & $ -\zeta+\beta_{1}$ & $\langle \hat{b}_2^\dagger \rangle$ & $\langle \bm{\hat{B}_4} \rangle $ & 1 & 1x1 & 1x1 & 1x1 & 1x1 \\
 \cline{2-9}
  & $ -\zeta-\beta_{1}$ & $\langle \hat{b}_3  \rangle$ & $\langle \bm{\hat{B}_5} \rangle $ & 1 & 1x1 & 1x1 & 1x1 & 1x1 \\
 \cline{2-9}
  & $ \zeta+\beta_{1}$ & $\langle \hat{b}_3^\dagger \rangle$ & $\langle \bm{\hat{B}_6} \rangle $ & 1 & 1x1 & 1x1 & 1x1 & 1x1 \\
 \cline{2-9}
  & 0 & $\langle \hat{b}_1 \rangle$ & $\langle \bm{\hat{B}_7} \rangle $ & 1 & 2x1 & 2x1 & 2x1 & 2x1 \\
 \cline{3-5}
  &  & $\langle \hat{b}_1^\dagger \rangle$ & $\langle \bm{\hat{B}_8} \rangle $ & 1 &  &   &  &\\
 \hline
 $2\gamma_+$ & $ -2\zeta\mp2\beta_{2}$ & $\langle \hat{b}_4^2 \rangle$, $\langle \hat{b}_5^{\dagger 2} \rangle$ & $ \langle \bm{\hat{B}_1^2}  \rangle $ & 1 & 1x4 & 1x4 & 1x3 & 1x3 \\
 & $ -2\zeta$ & $\langle \hat{b}_5^{\dagger } \hat{b}_4 \rangle$ & & 2 &  & &  &  \\
 \cline{2-9}
  &  $ 2\zeta\pm2\beta_{2}$ & $\langle \hat{b}_4^{\dagger 2} \rangle$, $\langle \hat{b}_5^{ 2} \rangle$  & $ \langle \bm{\hat{B}_2^2}  \rangle $  & 1 & 1x4 & 1x4 & 1x3 & 1x3 \\
  & $ 2\zeta$ & $\langle \hat{b}_4^{\dagger } \hat{b}_5 \rangle$ & & 2 &  &  &  &\\
  \cline{2-9}
  & $ 2\zeta-2\beta_{1}$ & $\langle \hat{b}_2^2 \rangle$ & $\langle \bm{\hat{B}_3^2} \rangle $ & 1 & 1x1 & 1x1 & 1x1 & 1x1  \\
  \cline{2-9}
  &  $ -2\zeta+2\beta_{1}$ & $\langle \hat{b}_2^{\dagger 2} \rangle$  & $\langle \bm{\hat{B}_4^2} \rangle $  & 1 & 1x1 & 1x1 & 1x1 & 1x1  \\
  \cline{2-9}
  &  $ -2\zeta-2\beta_{1}$ & $\langle \hat{b}_3^{ 2} \rangle$  & $\langle \bm{\hat{B}_5^2} \rangle $  & 1 & 1x1 & 1x1 & 1x1 & 1x1  \\
  \cline{2-9}
  &  $ 2\zeta+2\beta_{1}$ & $\langle \hat{b}_3^{\dagger 2} \rangle$  & $\langle \bm{\hat{B}_6^2} \rangle $  & 1 & 1x1 & 1x1 & 1x1 & 1x1  \\
  \cline{2-9}
  &  0 & $\langle \hat{b}_1^{2} \rangle$, $\langle \hat{b}_1^{\dagger 2} \rangle$  & $\langle \bm{\hat{B}_7^2} \rangle $,  $\langle \bm{\hat{B}_8^2} \rangle $ & 1 & 4x1 & 4x1 & 3x1 & 3x1  \\
  \cline{3-4}
  &   & $\langle \hat{b}_1^{\dagger} \hat{b}_1 \rangle$ & $\langle \bm{\hat{B}_8\hat{B}_7} \rangle $  & 2 &  &  &  & \\
  \cline{2-9}
  &  $ \pm2\beta_{2}$ & $\langle \hat{b}_4^{\dagger} \hat{b}_5^{\dagger} \rangle$, $\langle \hat{b}_4 \hat{b}_5 \rangle$ & $\langle \bm{\hat{B}_2\hat{B}_1} \rangle $  & 2 & 2x4 & 2x4 & 1x4 & 1x4 \\

  &  $ \beta_{2}-\beta_{2}$ & $\langle \hat{b}_4^{\dagger} \hat{b}_4 \rangle$, $\langle \hat{b}_5 \hat{b}_5^{\dagger} \rangle$ &  & 2 &  &  &  & \\
  \cline{2-9}
  &  $ -\beta_{1}\mp\beta_{2}$ & $\langle \hat{b}_4 \hat{b}_2 \rangle$,  $\langle \hat{b}_5^{\dagger} \hat{b}_2 \rangle$ & $\langle \bm{\hat{B}_1\hat{B}_3} \rangle $  & 2 & 4x1 & 4x1 & 2x1 & 2x1 \\
  \cline{2-9}
  &  $ 2\zeta-\beta_{1}\pm\beta_{2}$ & $\langle \hat{b}_4^{\dagger} \hat{b}_2 \rangle$,  $\langle \hat{b}_5 \hat{b}_2 \rangle$ & $\langle \bm{\hat{B}_2\hat{B}_3} \rangle $  & 2 & 4x1 & 4x1 & 2x1 & 2x1 \\
  \cline{2-9}
  &  $- 2\zeta +\beta_{1}\mp\beta_{2}$ & $\langle \hat{b}_2^{\dagger} \hat{b}_4 \rangle$,  $\langle \hat{b}_2^{\dagger} \hat{b}_5^{\dagger} \rangle$ & $\langle \bm{\hat{B}_4\hat{B}_1} \rangle $  & 2 & 4x1 &4x1  & 2x1 & 2x1 \\
  \cline{2-9}
  &  $\beta_{1}\pm\beta_{2}$ & $\langle \hat{b}_2^{\dagger} \hat{b}_4^{\dagger} \rangle$,  $\langle \hat{b}_2^{\dagger} \hat{b}_5 \rangle$ & $\langle \bm{\hat{B}_4\hat{B}_2} \rangle $  & 2 & 4x1 & 4x1 & 2x1 & 2x1 \\
  \cline{2-9}
  &  $ -2\zeta-\beta_{1}\mp\beta_{2}$ & $\langle \hat{b}_4 \hat{b}_3 \rangle$,  $\langle \hat{b}_5^{\dagger} \hat{b}_3 \rangle$ & $\langle \bm{\hat{B}_1\hat{B}_5} \rangle $  & 2 & 4x1 &4x1  & 2x1 & 2x1 \\
  \cline{2-9}
  &  $ -\beta_{1}\pm\beta_{2}$ & $\langle \hat{b}_4^{\dagger} \hat{b}_3 \rangle$,  $\langle \hat{b}_5 \hat{b}_3 \rangle$ & $\langle \bm{\hat{B}_2\hat{B}_5} \rangle $  & 2 & 4x1 & 4x1 & 2x1 & 2x1 \\
  \cline{2-9}
  &  $\beta_{1}\mp\beta_{2}$ & $\langle \hat{b}_3^{\dagger} \hat{b}_4 \rangle$,  $\langle \hat{b}_3^{\dagger} \hat{b}_5^{\dagger} \rangle$ & $\langle \bm{\hat{B}_6\hat{B}_1} \rangle $  & 2 & 4x1 &4x1  & 2x1 & 2x1 \\
  \cline{2-9}
  &  $ 2\zeta +\beta_{1}\pm\beta_{2}$ & $\langle \hat{b}_3^{\dagger} \hat{b}_4^{\dagger} \rangle$,  $\langle \hat{b}_3^{\dagger} \hat{b}_5 \rangle$ & $\langle \bm{\hat{B}_6\hat{B}_2} \rangle $  & 2 & 4x1 & 4x1 & 2x1 & 2x1 \\
  \cline{2-9}
  & $ -\zeta\mp\beta_{2}$ & $\langle \hat{b}_4 \hat{b}_1\rangle$, $\langle \hat{b}_5^\dagger \hat{b}_1 \rangle$ & $ \langle \bm{\hat{B}_7} \bm{\hat{B}_1} \rangle $ & 2 & 4x2 & 4x2 & 2x2 & 2x2 \\
  \cline{3-4}
  &  & $\langle \hat{b}_4 \hat{b}_1^\dagger\rangle$, $\langle \hat{b}_5^\dagger \hat{b}_1^\dagger \rangle$ & $ \langle \bm{\hat{B}_8} \bm{\hat{B}_1} \rangle $ & 2 &  &  &  &  \\
  \cline{2-9}
  & $ \zeta\pm\beta_{2}$ & $\langle \hat{b}_4^\dagger \hat{b}_1\rangle$, $\langle \hat{b}_5 \hat{b}_1 \rangle$ & $ \langle \bm{\hat{B}_7} \bm{\hat{B}_2} \rangle $ & 2 & 4x2 & 4x2 & 2x2 & 2x2 \\
  \cline{3-4}
  &  & $\langle \hat{b}_4^\dagger \hat{b}_1^\dagger\rangle$, $\langle \hat{b}_5 \hat{b}_1^\dagger \rangle$ & $ \langle \bm{\hat{B}_8} \bm{\hat{B}_2} \rangle $ & 2 &  &  &  &  \\
  \cline{2-9}
  & $ \beta_{1}-\beta_{1}$ & $\langle \hat{b}_2^\dagger \hat{b}_2 \rangle$ & $\langle \bm{\hat{B}_4\hat{B}_3} \rangle $   & 2 & 2x1 & 2x1 & 1x1 & 1x1 \\
  \cline{2-9}
  & $ -2\beta_{1}$ & $\langle \hat{b}_3 \hat{b}_2 \rangle$ & $\langle \bm{\hat{B}_5\hat{B}_3} \rangle $   & 2 & 2x1 & 2x1 & 1x1 & 1x1 \\
  \cline{2-9}
  & $ 2 \zeta$ & $\langle \hat{b}_3^\dagger \hat{b}_2 \rangle$ & $\langle \bm{\hat{B}_6\hat{B}_3} \rangle $   & 2 & 2x1 & 2x1 & 1x1 & 1x1 \\
  \cline{2-9}
  & $ \zeta-\beta_{1}$ & $\langle \hat{b}_1 \hat{b}_2\rangle$ & $ \langle \bm{\hat{B}_7} \bm{\hat{B}_3} \rangle $ & 2 & 4x1 & 4x1 & 2x1 & 2x1 \\
  \cline{3-4}
  &  & $\langle \hat{b}_1^\dagger \hat{b}_2\rangle$ & $ \langle \bm{\hat{B}_8} \bm{\hat{B}_3} \rangle $ & 2 &  &  &  &  \\
  \cline{2-9}
  & $ -2\zeta$ & $\langle \hat{b}_2^\dagger \hat{b}_3 \rangle$ & $\langle \bm{\hat{B}_4\hat{B}_5} \rangle $   & 2 & 2x1 & 2x1 & 1x1 & 1x1 \\
  \cline{2-9}
  & $ 2 \beta_1$ & $\langle \hat{b}_2^\dagger \hat{b}_3^\dagger \rangle$ & $\langle \bm{\hat{B}_4\hat{B}_6} \rangle $   & 2 & 2x1 & 2x1 & 1x1 & 1x1 \\
  \cline{2-9}
  & $ -\zeta-\beta_{1}$ & $\langle \hat{b}_2^\dagger \hat{b}_1\rangle$ & $ \langle \bm{\hat{B}_4} \bm{\hat{B}_7} \rangle $ & 2 & 4x1 & 4x1 & 2x1 & 2x1 \\
  \cline{3-4}
  &  & $\langle \hat{b}_2^\dagger \hat{b}_1^\dagger \rangle$ & $ \langle \bm{\hat{B}_4} \bm{\hat{B}_8} \rangle $ & 2 &  &  &  &  \\
  \cline{2-9}
  & $ \beta_1- \beta_1$ & $\langle \hat{b}_3^\dagger \hat{b}_3 \rangle$ & $\langle \bm{\hat{B}_6\hat{B}_5} \rangle $   & 2 & 2x1 & 2x1 & 1x1 & 1x1 \\
  \cline{2-9}
  & $ -\zeta-\beta_{1}$ & $\langle \hat{b}_1 \hat{b}_3\rangle$ & $ \langle \bm{\hat{B}_7} \bm{\hat{B}_5} \rangle $ & 2 & 4x1 & 4x1 & 2x1 & 2x1 \\
  \cline{3-4}
  &  & $\langle \hat{b}_1^\dagger \hat{b}_3\rangle$ & $ \langle \bm{\hat{B}_8} \bm{\hat{B}_5} \rangle $ & 2 &  &  &  &  \\
  \cline{2-9}
  & $ \zeta+\beta_{1}$ & $\langle \hat{b}_3^\dagger \hat{b}_1 \rangle$ & $ \langle \bm{\hat{B}_6} \bm{\hat{B}_7} \rangle $ & 2 & 4x1 & 4x1 & 2x1 & 2x1 \\
  \cline{3-4}
  &  & $\langle \hat{b}_3^\dagger \hat{b}_1^\dagger \rangle$ & $ \langle \bm{\hat{B}_6} \bm{\hat{B}_8} \rangle $ & 2 &  &  &  &  \\
  \hline
 \end{tabular}
 \vspace{-2mm}
  \caption{Real and imaginary parts of the complex eigenfrequencies $
  \bm{\Lambda_{j}^{\rm r} } - i
  \bm{\Lambda_{j}^{\rm i} } $ of the matrix $ \bm{M^{(5)}_{\rm p}} $, given in Eq.~(\ref{51}),
   for the five-mode pyramid bosonic system
   derived from the equations for the
   FOMs up to second order assuming $ \beta_2 = 0 $.
   If $ \beta_1 = 0 $ instead of $ \beta_2 = 0 $, simple relabelling of $ \hat{b}_j $,
   $ \hat{b}_j^\dagger $ operators provides the corresponding table.
   {We have $ \bm{\hat{B}_1} \equiv [\hat{b}_4,\hat{b}_5^\dagger] $,
    $ \bm{\hat{B}_2} = \bm{\hat{B}_1}^\dagger $,  $ \bm{\hat{B}_3} \equiv
   [\hat{b}_2] $, $ \bm{\hat{B}_4} = \bm{\hat{B}_3}^\dagger $, $ \bm{\hat{B}_5} \equiv
   [\hat{b}_3] $, $ \bm{\hat{B}_6} = \bm{\hat{B}_5}^\dagger $, $ \bm{\hat{B}_7} \equiv
   [\hat{b}_1] $, $ \bm{\hat{B}_8} = \bm{\hat{B}_7}^\dagger $,
   and more details are given in the caption to Tab.~\ref{tab1}.} }
 \label{tab6}
\end{table*}

Finally, for the five-mode bosonic system in the pyramid
configuration, described by the dynamical matrix ${\bm M}_{\rm
p}^{(5)}$ given in Eq.~(\ref{51}) with damping and/or
amplification rates satisfying the conditions in Eqs.~(\ref{52})
and (\ref{53}), the QEPs and QHPs identified in the dynamics of
first- and second-order FOMs are systematically summarized in
Tab.~\ref{tab6}. Genuine QEPs with second-, third-, and
fourth-order EDs appear in the spectral structure of second-order
FOMs, though with relatively low DDs. This reflects the limited
presence of DDs in the first-order FOM spectra, where only one
eigenfrequency exhibits a second-order DD.

\bibliographystyle{quantum}
\bibliography{thapliyal}

\end{document}